\documentclass[apj]{emulateapj}
\usepackage{xspace}
\usepackage{amsmath}
\usepackage{epsfig}
\usepackage{graphicx}
\usepackage{rotating}
\usepackage{afterpage}

\def\nW{$\mathrm{nWm^{-2}sr^{-1}}$}
\def\mic{$\mu {\rm m}$}
\def\NLF{$233$ }
\def\HFE{{\small HFE}}
\def\LFE{{\small LFE}}

\def\bUV{{\small $\widetilde{UV}$}}
\def\bU{{\small $\widetilde{U}$}}
\def\bB{{\small $\widetilde{B}$}}
\def\bV{{\small $\widetilde{V}$}}
\def\bR{{\small $\widetilde{R}$}}
\def\bI{{\small $\widetilde{I}$}}
\def\bz{{\small $\widetilde{z}$}}
\def\bJ{{\small $\widetilde{J}$}}
\def\bH{{\small $\widetilde{H}$}}
\def\bK{{\small $\widetilde{K}$}}
\def\bL{{\small $\widetilde{L}$}}
\def\bM{{\small $\widetilde{M}$}}

\begin{document}

\pagestyle{myheadings} \markright{DRAFT: \today\hfill}

\title{Reconstructing the Near-IR Background Fluctuations from known Galaxy Populations
using Multiband Measurements of Luminosity Functions}

\author{Kari Helgason$^{1,2}$, Massimo Ricotti$^1$, Alexander Kashlinsky$^2$ }
\affil{$^1$Department of Astronomy, University of
Maryland, College Park, MD 20742, USA; kari@astro.umd.edu}
\affil{$^2$SSAI and Observational Cosmology Laboratory, Code 665, NASA Goddard Space
Flight Center, Greenbelt MD 20771}

\begin{abstract}

We model fluctuations in the Cosmic Infrared Background (CIB) arising from known galaxy populations using \NLF measured UV, optical and NIR luminosity functions (LF) from a variety of surveys spanning a wide range of redshifts. We compare best-fit Schechter parameters across the literature and find clear indication of evolution with redshift. Providing fitting formulae for the multi-band evolution of the LFs out to z$\sim$5, we calculate the total emission redshifted into the near-IR bands in the observer frame and recover the observed optical and near-IR galaxy counts to a good accuracy. Our empirical approach, in conjunction with a halo model describing the clustering of galaxies, allows us to compute the fluctuations of the unresolved CIB and compare the models to current measurements.  We find that fluctuations from known galaxy populations are unable to account for the large scale CIB clustering signal seen by {\it Spitzer}/IRAC and {\it AKARI}/IRC and continue to diverge out to larger angular scales. This holds true even if the LFs are extrapolated out to faint magnitudes with a steep faint-end slope all the way to $z$=8. We also show that removing resolved sources to progressively fainter magnitude limits, isolates CIB fluctuations to increasingly higher redshifts. Our empirical approach suggests that known galaxy populations are not responsible for the bulk of the fluctuation signal seen in the measurements and favors a very faint population of highly clustered sources.

\end{abstract}



\section{Introduction}


Cosmic infrared background (CIB) includes contributions from emissions over the entire history of the Universe, including from objects inaccessible to the current telescopic studies.
Several direct measurements of the total mean levels of the CIB using the wide-beam DIRBE and IRTS instruments claim a significant excess mean flux over the contribution of known galaxies in the near-IR \citep{Dwek98,Gorjian00,Wright00,Cambresy01,Matsumoto05}; also see review by \cite{Kashreview}. The entire  excess emission over that from known galaxy populations ($\simeq$ 30 \nW\ in 1-4\mic, \cite{Kashreview}) was originally theorized to come from primordial PopIII stars \citep{Santos02,Salvaterra&Ferrara03} but this interpretation has been challenged on several grounds since the claimed levels require uncomfortable levels of star formation efficiency \citep{Madau&Silk05,Salvaterra&Ferrara06}. It is possible that much of the excess flux seen may be due to inaccurate removal of bright zodiacal emission in the foreground \citep{Dwek&Arendt05,Mattila06}. Furthermore, the extragalactic background light (EBL) is a fundamental source of opacity for high energy photons and the $\gamma$-ray attenuation seen in blazar spectra favors low levels of NIR background light \citep[e.g.][]{Aharonian06,Mazin&Raue07}.

An alternative way to study the CIB, much less sensitive to foreground removal, is to measure background anisotropies after subtracting resolved galaxies down to faint magnitudes \citep{Kashlinsky96}. Detections of spatial structure in the CIB were initially based on datasets from {\it COBE}/DIRBE \citep{Kashlinsky00}, the {\it IRTS} \citep{Matsumoto00} and {\it 2MASS} \citep{Kashlinsky02,Odenwald03}. More recently, \citet{KAMM1,KAMM2} using deep exposures from {\it Spitzer}/IRAC (3.6-8.0\mic) found significant fluctuations after subtracting galaxies down to $m_{AB}$$\approx$25. The level of these fluctuations, $\sim$0.1 \nW\ at arcminute scales, imply an isotropic CIB flux as low as $\sim$1 \nW\ from the remaining unresolved sources in the IRAC bands \citep{KAMM3}. \citet{Thompson07a} analysis constrained CIB at 1.4-1.8\mic\ using {\it HST}/NICMOS observations and \cite{Matsumoto11} measure fluctuations on arcminute scales in the $2.4$-$4.1$\mic\ range using the {\it AKARI} satellite. After this paper was submitted, \citet{Kashlinsky12} have measured the {\it Spitzer}/IRAC out to $\lesssim$1$^\circ$ using more extensive datasets from the {\it Spitzer Extended Deep Survey} (SEDS), confirming earlier results and extending the fluctuation measurement to much larger angular scales. All the present measurements of CIB-fluctuations are consistent with an extragalactic origin, necessitating an associated unresolved component in the CIB. This component likely requires only a fraction of the CIB excess, which is below limits imposed by $\gamma$-ray photon absorption \citep{KAMM3,Kashlinsky&Band07,Arendt10}.

There seems to be an emerging consensus that the extragalactic clustering signal is real, but the nature of the sources producing it is still a subject of debate. Plausible candidates for the bulk of the CIB are evolving stellar populations in galaxies, although accreting black holes at high-$z$ can also contribute \citep[e.g.,][]{RicottiO04}. Both \citet{KAMM1} and \citet{Matsumoto11} argue that that the clustering is consistent with "first stars"  era objects whereas \citet{Cooray07,Chary08} have posited that the signal originates mostly in the clustering of faint galaxies at redshifts $z$$\sim$1-3. Understanding the expected levels of fluctuations from known galactic populations is possible following the establishment of the standard cosmological model for structure formation, the concordance $\Lambda$CDM \citep{Komatsu11}. In order to compute the levels of source-subtracted CIB fluctuations remaining in the {\it Spitzer} data, \citet{Sullivan07} used a halo model combined with conditional luminosity functions and compared it to measurements at 3.6\mic. Their claim is that the fluctuations detected by \citet{KAMM1} can be explained by ordinary galaxies just beyond the detection threshold of {\it Spitzer}/IRAC, although this claim appears to contradict the results of their analysis shown in their Fig.~8.

\cite{Kashreview} discusses the importance of the shape of the emission history for the resulting fluctuations demonstrating how brief episodes of light production can lead to enhanced fluctuations. In this paper, we construct the entire history of light production produced by known galaxy populations using a novel empirical approach that relies exclusively on observations. We use a compilation of galaxy luminosity functions (LF) in the literature to populate the observed lightcone with galaxies down to faint magnitudes. The many galaxy surveys conducted in recent years provide a wealth of data in multiple bands and cover a wide range of redshifts. Individually, LFs only probe specific rest-frame wavelengths for a limited range of redshifts, while together we can use them to infer the source distribution composing the background light in the 0.1-5.0\mic\ range. Our only theoretical assumptions concern the clustering properties of the unresolved sources which are modeled according to the well-established concordance $\Lambda$CDM model (see Section~\ref{sec:fluctuations}). We refer to \citet{Johnston11} for a good review on the properties of luminosity functions and how they are measured.

Modeling the underlying populations of the EBL has been attempted using various mixtures of theory and observations. Backward evolution scenarios take the present galaxy populations and extrapolate them to higher redshift \citep[e.g.,][]{Jimenez&Kashlinsky99,Franceschini08}, while forward evolution follows dark matter merger trees starting from the cosmological initial conditions, using semi-analytical models of galaxy formation \citep{Gilmore10,Guo11}. \citet{Dominguez11} use directly the measured K-band LFs out to z=4 from \cite{Cirasuolo10}, combined with best-fit SEDs of multi-wavelength galaxy data (AEGIS) to empirically derive the overall EBL spectrum. We however, present an alternative empirical approach by examining the best-fit Schechter parameters \citep{Schechter76} of \NLF LFs covering the UV, optical and near-IR out to redshifts z$\sim$3-8. We provide empirical fitting formulae describing the smooth evolution of multi-band LFs with redshift, and construct lightcones containing all populations seen in the near-IR bands, selected at each redshift such that $\lambda_{NIR}^{obs} = (1+z)\lambda^{rest}$.

This paper is organized as follows. Section~\ref{sec:lf_data} describes the data used and in Section~\ref{sec:poplight} we explain the modeling in detail. In Section~\ref{sec:nc}, we calculate both galaxy number counts and the EBL in the near-IR bands ($JHKLM$) and compare with existing data. In Section~\ref{sec:fluctuations} we analyze the source-subtracted CIB-fluctuations implied by our empirical reconstruction and compare with previous work. We discuss the implications of our findings in Section~\ref{sec:discussion}.
Throughout this paper we adopt the concordance $\Lambda$CDM comology with $\Omega_m$=0.3, $\Omega_{\Lambda}$=0.7 and $H_0$=70 $\mathrm{km \! \cdot \! s^{-1}\! \cdot\! Mpc^{-1}}$. All magnitudes are in the AB system unless stated otherwise \citep{Oke&Gunn83}.

\section{Measurements of the Galaxy Luminosity Function} \label{sec:lf_data}

\begin{figure}[b]
      \includegraphics[width=.47\textwidth]{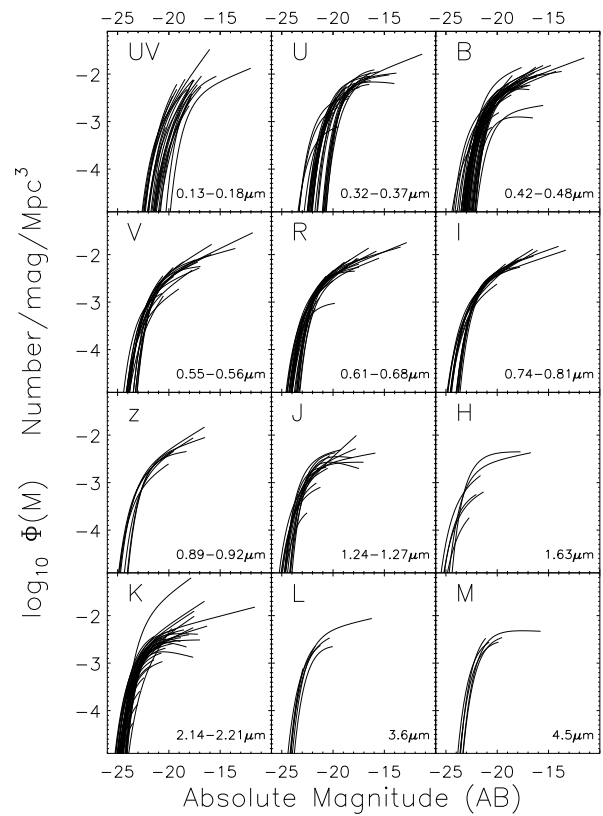}
      \caption{\footnotesize All \NLF luminosity functions used in our analysis in Schechter parametrization (see original references in Table~\ref{lf_data}). The wavelength bins are listed in the panels (lower right) and their effective wavelengths are listed in Table~\ref{tab:bands} along with other properties. The LFs shown have a range of redshifts. }
      \label{fig_lf}
\end{figure}
\begin{sidewaystable*}[h]
\begin{center}
\caption{Measurements of the Luminosity Function}
\vspace{200pt}
\begin{tabular}{ l c c c c c c }
 Reference & Rest-frame band & Redshift & Sample & Selection & Survey Catalog / Field  &  Symbol / Color$^b$    \\
           &                 &$ z $ & $N_{gal}$ & $m_{lim} (AB)$ &       &    \\

\hline
\hline

\cite{Arnouts05}        & 1500\AA  & 0.2-1.2    & 1039  &    NUV$<$24.5   & GALEX/VVDS  &       green triangles(up)             \\
                        &          & 1.75-3.4   &       &      F450\&F606$<$27 & HDF    &                      \\
\cite{Wyder05}          & {\footnotesize $\mathrm{NUV,FUV}$}  & 0.055      & 896,1124 & $m_{UV}\!<\!20$   & GALEX/2dF &   blue circles  \\
\cite{Oesch10}       & 1500\AA  & 0.5-2.5    & 284-403  &    $\lesssim$26   & HST ERS &           yellow circles          \\
\cite{Oesch12}       & 1500\AA  & $\sim$8    & 70  &    $H$$<$27.5   & CANDLES/HUDF09/ERS  &           pink triangles(up)   \\
\cite{Reddy08}         &  1700\AA & 1.9-3.4   & $\sim$15,000  & $\mathcal{R}$$<$25.5     & $^a$ &        blue crosses      \\
\cite{Yoshida06}        & 1500\AA         & $\sim$4,5   &  3808,539   & $\lesssim$26-27   & Subaru Deep Field &    blue squares        \\
\cite{McLure09}       & 1500\AA  & $\sim$5,6    & $\sim$1500  &    $z^\prime$$\lesssim$26   & SXDS/UKIDSS  &   purple squres       \\
\cite{Ouchi09}       & 1500\AA  & 7    & 22  &    $\lesssim$26   & SDF/GOODS-N \\
\cite{Bouwens07}        & 1600\AA,1350\AA & $\sim$4,5,6 &  4671,1416,627 & $\lesssim$29 &  HUDF/GOODS   &       violet triangles(down)     \\
\cite{Bouwens11}        & 1600\AA,1750\AA & $\sim$7,8   &  73,59      & $\lesssim$26-29.4 &  HUDF09    &        orange diamonds         \\

\cite{Gabasch04}        & $u^\prime g^\prime$ & 0.45-5 & 5558 & $I\! <\! 26.8$ & FORS Deep Field &   Green triangles(down)          \\
\cite{Baldry05}         & $^{0.1}u$        & $<$0.3      &   43223     & u$<$20.5         & SDSS     &   red squares       \\
\cite{Faber07}          & $B$  &  0.2-1.2 & $\sim$34000 & $R\lesssim24$ & DEEP2/COMBO-17        &      tan squares        \\
\cite{Norberg02}        &$b_j$            & $<$0.2       & 110500     &  $<$19.45 & 2dFGRS       &     purple squares            \\
\cite{Blanton03}        & $^{0.1}ugriz$ & 0.1 & 147986 & $<$16.5-18.3 & SDSS                 &   blue plus \\
\cite{Montero-Dorta09}  &$^{0.1}ugriz$ & $\lesssim$0.2 & 947053 & $<$17-19 & SDSS           &   green crosses      \\
\cite{Loveday12}        &$^{0.1}ugriz$ & 0.002-0.5 & 8647-12860  & $r\!<\!19.8$ & GAMA      &    yellow squares       \\

\cite{Ilbert05}         & $UBVRI$ & 0.05-2.0 & 11034 & $I\!<\!24$ & VIMOS-VLT Deep Survey    &   pink triangles(up)           \\
\cite{Gabasch06}        & $i^\prime z^\prime r^\prime$  & 0.45-3.8 & 5558 & $I\!<\!26.8$ & FDF      &     green circles         \\
\cite{Marchesini07}     & $BVR$   & 2.0-3.5  & 989   & $K_s\! \lesssim \! 25$ & MUSYC/FIRES/GOODS/EIS     &    orange circles   \\
\cite{Marchesini12}     & $V$   & 0.4-4.0  & 19403 & $H$$<$27.8,$K$$<$25.6 & $^a$               &     blue triangles(up)    \\

\cite{Hill10}           & $ugriz$       & 0.0033-0.1  &  2437-3267  & $<$18-21  & MGC/UKIDSS/SDSS    &     purple diamonds     \\
                        & $YJHK$            &             &  1589-1798  & $<$17.5-18 &               &         \\
\cite{Dahlen05}         & $UBR$   & 0.1-2  &  18381 & $R\!<\!24.5$   &  GOODS-HST/CTIO/ESO         &     dark green diamonds       \\
                        & $J$   &   0.1-1  & 2768   & $K_s\! <\! 23.2$ &   & \\
\cite{Jones06}          &$b_jr_f$      &  $<$0.2      & 138226  & $b_jr_f\! < \! 15.6,16.8$ & 6dFGS/2MASS    &  dark red plus      \\
                        &$JHK$        &               &         & $JHK \! < \! 14.7$ & /SuperCOSMOS       &         \\
\cite{Bell03}           & $ugriz$    & $<0.1$ &    22679 & $r \! < \!17.5$ & SDSS                  &        orange circles                \\
                        & $K$        &        &    6282  & $K \! < \! 15.5$   & 2MASS         &            \\
\cite{Kashikawa03}      & $BK^\prime$       & 0.6-3.5  &  439 &   $K^\prime \! < \! 24$ & Subaru Deep Survey     &    red crosses      \\
\cite{Stefanon11}       & $JH$       & 1.5-3.5  &  3496 &  $K_s \! < \! 22.7$-$25.5$ & MUSYC/FIRES/FIREWORKS    &     green squares        \\
\cite{Pozzetti03}       & $JK_s$       & 0.2-1.3    &  489   &  $K_s \! < \!20$ & K20 Survey                    &    tan plus    \\
\cite{Feulner03}        & $JK^\prime$    & 0.1-0.6  & 500 & $K^\prime \! < \! 19.4$-$20.9$ &  MUNICS             &    yellow crosses        \\
\cite{Eke05}            & $JK_s$     & 0.01-0.12   & 16922,15664 & $JK_s\! \lesssim \! 15.5$ & 2dFGRS/2MASS     &          violet diamonds    \\
\cite{Cole01}           & $JK_s$     & 0.005-0.2   & 7081,5683 &  $JK_s \! \lesssim \! 15.5$ & 2dFGRS/2MASS      &         blue squares      \\
\cite{Smith09}          & $K$        & 0.01-0.3 & 40111 &  $K \! < \! 17.9$,$r \! < \! 17.6$  & UKIDSS-LAS/SDSS    &       red triangle(up)        \\
\cite{Saracco06}        & $K_s$      & 0.001-4  & 285 &    $Ks \! < \! 24.9 $  & HDFS/FIRES      &        blue triangles(down)     \\
\cite{Kochanek01}       & $K_s$      & 0.003-0.03  & 4192      &  $K_{20} \! < \!13.35$ & 2MASS/CfA2/UZC     &    magenta circles             \\
\cite{Huang03}          & $K$        & 0.001-0.57 &  1056  &   $K \! < \! 15$ & 2dF/AAO          &        violet diamonds          \\
\cite{Arnouts07}        & $K$     & 0.2-2  & 21200 & $m_{3.6mic} \! < \! 21.5$ & SWIRE/VVDS        &      dark green squares           \\
                        &           &            & &                          & /UKIDSS/CFHTLS      &            \\
\cite{Cirasuolo10}      & $K$ &  0.2-4     & $\sim$50000 & $K \! < \! 23$ & UKIDSS/SXDS               &   orange plus        \\
\cite{Babbedge06}       & $L_{3.6\mu m}M_{4.5\mu m}$ & 0.01-0.6 & 34281 & $ \! < \! 20.2$ & SWIRE/INT WFS    &   blue crosses   \\
\cite{Dai09}            & $L_{3.6\mu m}M_{4.5\mu m}$ & 0.01-0.6 &  4905,5847 & $LM \! < \! 19,I \! < \! 20.4$ & IRAC-SS/AGES    &  dark red circles        \\

\hline
\end{tabular}
\tablecomments{The measured LF are shown in Figure~\ref{fig_lf} and all Schechter parameters are displayed in Figure~\ref{fig_params}. The compile database of the Schechter parameters is available upon request. \\
$^a$Data taken from multiple surveys/fields \\
$^b$The symbols and color of the corresponding data points in Figure \ref{fig_rhoL} and \ref{fig_params}   }
\hfill
\label{lf_data}
\end{center}
\end{sidewaystable*}

The total emission seen in the near-IR bands ({\small $JHKLM$}) depends on the contribution of local near-IR galaxies as well as redshifted light radiated at shorter rest-frame wavelengths. To quantify the present day background produced by galaxies, we have utilized measurements of luminosity functions probing all {\em rest-frame} wavelengths in the interval 0.1$<\! \lambda \!<$5.0\mic\ anywhere in the redshift cone. This results in a compilation of \NLF LFs from a large variety of surveys which we list in Table~\ref{lf_data}. Our approach does not depend on stellar population synthesis models \citep[e.g.,][]{Bruzual&Charlot03} and we do not need to make an assumption for the IMF. Rather, in this method we predict the levels of CIB fluctuations directly from the available data, assuming only i) standard $\Lambda$CDM model of structure formation and ii), the validity of a Schechter-type LF after fitting its parameters to the data. All the LFs we use have been characterized by a Schechter function \citep{Schechter76},
\begin{equation} \label{eqn:Schechter}
  \begin{split}
  \phi(M)dM = 0.4 \ln{(10)}\phi^{\star} & \left(10^{0.4(M^{\star}-M)}\right)^{\alpha + 1} \\
   & \times \exp{(-10^{0.4(M^{\star}-M)})}dM,
  \end{split}
\end{equation}
determined by the normalization, $\phi^{\star}$ , characteristic absolute magnitude, $M^{\star}$ and the faint-end slope, $\alpha$. By integrating Equation~(\ref{eqn:Schechter}), the luminosity density can be shown to be  $\mathcal{L}=\phi^\star L^\star \Gamma(\alpha+2)$, where $L^\star$ is the characteristic luminosity and $\Gamma(x)$ is the Gamma function. All the Schechter LFs used are shown in Figure~\ref{fig_lf}.

The Schechter LF is usually found to fit the data fairly accurately but deviations are seen, in particular when fitting a wide range of luminosities. At low-$z$ for example, \citet{Jones06} find that the shape does not fit the sharp downturn seen at $M^\star$ and both \citet{Blanton03} and \citet{Montero-Dorta09} find an excess of bright galaxies in the blue SDSS bands. There are also hints of an upturn in the local LF at faint magnitudes where the Schechter fit does a poor job \citep[e.g.,][]{Blanton05}. We address this faint-end issue in Section~\ref{sec:faintend}, but note that sources at the bright end are efficiently removed from the maps in CIB fluctuations studies. At longer wavelengths ($>$5\mic), a double power-law is found to provide a more adequate fit than the Schechter function \citep{Babbedge06,Magnelli11}. The mutual consistency of measurements is a primary concern when comparing LFs in the literature. Inconsistencies can be caused by field-to-field variations, photometric system, k-corrections, type of LF-estimator, survey depth and completeness, redshift binning, sample statistics, error estimates, etc. These undoubtedly account for differences in shape and amplitude of the measured LF (see Figure~\ref{fig_params}). We include a discussion of common issues in Appendix~\ref{sec:consistency}, but these do not affect our results because we let all measurements collectively contribute to our derived LF (see Section~\ref{sec:poplight}).

To directly compare flux measurements at different wavelengths, we have adopted the $AB$ magnitude system which conveniently relates the apparent magnitude, $m_{AB}$, to the specific flux, $f_\nu$, via
\begin{equation} \label{eqn:AB}
f_\nu= 10^{-0.4(m_{AB}-23.9)} \mu {\rm Jy},
\end{equation}
(1 Jy = 10$^{-26}$Wm$^{-2}$Hz$^{-1}$).
Where system conversions are not explicitly given by the authors (Table~\ref{lf_data}) we have made use of the calculations available at http://mips.as.arizona.edu/$\sim$cnaw/sun.html. With all magnitudes converted to $AB$ we do not distinguish between magnitudes of different filter- and photometric variations, e.g. {\small Johnson} $U$ and {\small SDSS} $u$, apart from their center frequencies.

\begin{figure}
      \includegraphics[width=.49\textwidth]{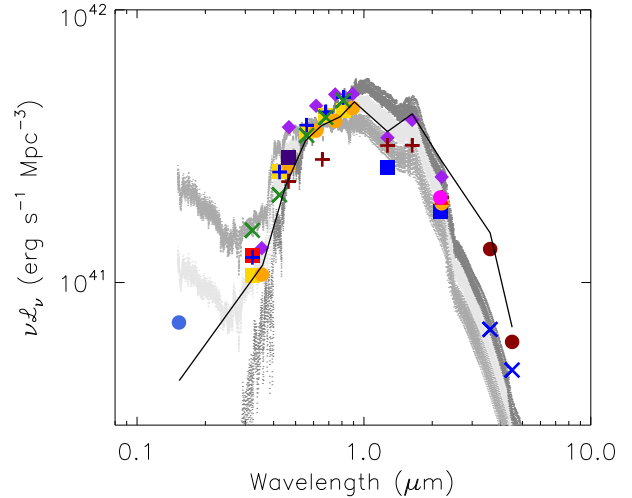}
      \caption{The local luminosity density according to all available LF measurements at z$<$0.12 in Table~\ref{lf_data}, with symbols/colors indicated in the same Table. To avoid overcrowding the region of interest we omit error bars. The solid line shows the luminosity density in our fiducial bands as implied by our fits in Figure~\ref{fig_params}. The sets of gray lines show the contribution from galaxies of different metallicities and ages from synthetic galaxy SED spectra shown in Fig. 14 of \citep{Kashreview}. The bottom-gray curves show the early type stellar populations, the upper-dark show late type populations and middle-light lines show the average of the two contributions. }
      \label{fig_rhoL}
\end{figure}

\section{Populating the Lightcone with known Galaxy Populations} \label{sec:poplight}

This section outlines the step-by-step approach leading to the quantification of the galaxy distribution seen on the sky. Using the data in Table~\ref{lf_data}, we populate the evolving lightcone by placing the rest-frame galaxy distribution at a distance such that the associated emission is shifted into the near-IR bands in the observer frame, defined by $\lambda_{NIR}/(1+z)$. Initially, we bin the LFs according to their rest-frame wavelength in fiducial bands which we call \bUV, \bU, \bB, \bV, \bR, \bI, \bz, \bJ, \bH, \bK, \bL\ and \bM\ (see Table~\ref{tab:bands}). For example, measurements in rest-frame {\small SDSS} $g^\prime$, {\small Johnson} $B$ and 2dF $b_j$ are binned together in our \bB-band despite having an offset in center wavelength of about 0.03\mic. The largest offset occurs in our \bI-bin where the centers of {\small SDSS} $i$ and {\small Johnson} $I$ is 0.063\mic. The uncertainty associated with the redshift of the population usually dominates these offsets so we do not correct for them. The centers of our fiducial bands, $\lambda_{eff}$, are taken to be the mean rest-frame wavelength of all measurements in the bin (see Table~\ref{tab:bands}).

\begin{sidewaysfigure*}

\vspace{250pt}
\hspace{-70pt}
      \includegraphics[width=1.2\textwidth]{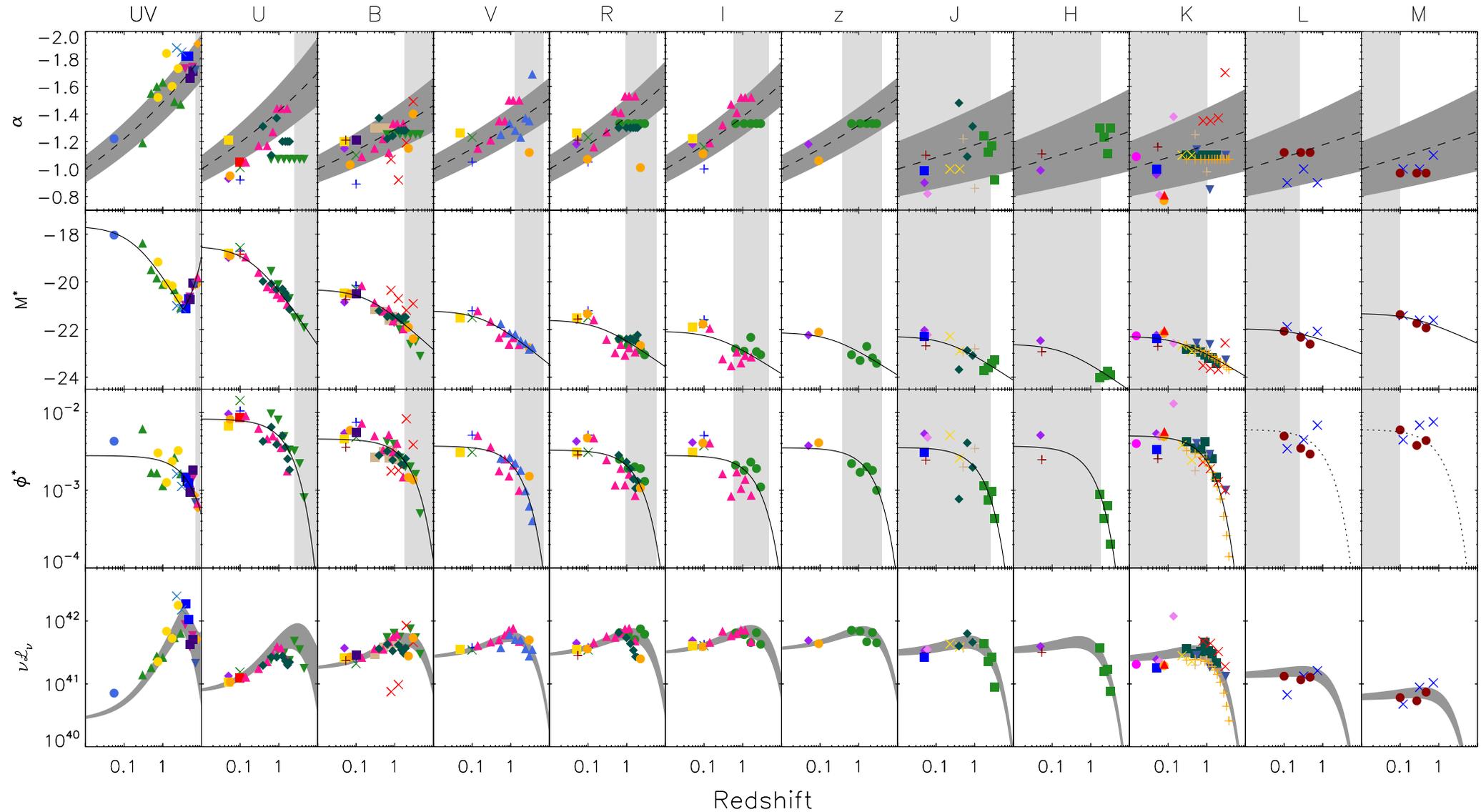}
      \caption{ The measured Schechter parameters $\alpha$,$M^\star$,$\phi^\star$ from the studies in Table~\ref{lf_data} including the luminosity density, $\mathcal{L}_\nu$=$\phi^\star L^\star \Gamma(\alpha+2)$, as a function of redshift. The different sybmols/colors are listed along with the corresponding references in Table~\ref{lf_data}. We have omitted error bars for the sake of clarity. The solid curves show the evolutionary fits according to Equations~(\ref{Msfit})-(\ref{psfit}) with the best-fit parameters listed in Table~\ref{tab:bands}. We have modified $M^\star_{UV}$ to follow the fitting functions of \citet{Bouwens11} at z$>$3.5 to better match the turnover seen. We note that our fits are only empirically supported for z$\lesssim$4, beyond which we extrapolate. The dashed curves in the $\alpha$ panels shows the evolution assumed in our default model whereas the dark shades areas encompass the range bracketed by our {\it high faint-end} (\HFE) and {\it low faint-end} (\LFE) models. These ranges are ultimately constrained by the observed galaxy counts (see Section~\ref{sec:faintend}). The shaded areas in the bottom row ($\nu\mathcal{L}_\nu$) is the evolving quantity $\phi(z)^\star \nu L_\nu(z)^\star \Gamma(\alpha(z)+2)$ corresponding to this allowed range in $\alpha(z)$. The dotted curves in $\phi^\star$ in \bL\ and \bM-bands are not fits to the data but are instead assumed to have the same form as the \bK-band fits. The light gray shaded areas correspond to the redshift regions for which the rest-frame emission redshifts into the observed NIR wavelengths of interest, defined to encompass the 1.25-4.5\mic\ range. We are most concerned with the goodness of fit in these regimes. All the data-points assume $h$=0.7. }
      \label{fig_params}

\end{sidewaysfigure*}

By placing the entire population of each LF at the median redshift of the sample, $z_{med}$, we examine the evolution of the individual Schechter parameters ($\alpha$,$M^\star$,$\phi^\star$) in our fiducial bands. In the cases where $z_{med}$ is not explicitly given by the authors, we choose the midpoint of the redshift bin of the LF measurement. The distances of the galaxies composing the LF is the dominant uncertainty in the resulting counts on the sky and we have therefore examined the effects of placing the LF at the opposite boundaries of the redshift bin (the resulting counts differ by less than a factor of two at the two extremes (see Section~\ref{sec:nc})). Figure~\ref{fig_params} shows the Schechter parameters as a function of redshift from 0.15-4.5\mic. Across the spectrum, we see clear indication of evolution in $M^\star$ and $\phi^\star$ and in some cases also in the poorly measured $\alpha$.

Over time, galaxy populations evolve both in brightness and abundance. As small systems merge to form more massive ones, we expect a net increase in the number of bright and massive galaxies with time accompanied by a decrease in fainter ones. This is encoded in the evolution of $\phi^\star$ (the number density of $L^\star$ systems), which we expect to increase with time whereas the faint-end slope, $\alpha$ should consequently flatten. The difference of the LF among rest-frame bands reflects the tendency of galaxies of different types being preferentially bright/faint at a given wavelength. The decomposition of the LF into red/blue galaxies typically shows an early-type population of individually bright galaxies with a diminishing faint-end whereas a the late-type population is composed of a rising number of faint galaxies \citep[e.g.,][]{Faber07}. The characteristic luminosity, $L^\star$ therefore depends heavily on the mixture of spectral types at any given epoch. Much work has been devoted to the K-band LF where the stellar mass-to-light ratio is relatively stable and it can thus be used as an indicator for the stellar mass function \citep[e.g.,][]{Cole01}. It is therefore natural to expect $M_K^\star$ to brighten with cosmic time as more mass becomes locked up in low-mass stars. In the red/NIR bands, the luminosity evolution is typically $\Delta M^\star \! \simeq \! 0.5-1.0$ between redshift 0.1 and 1 whereas it is much stronger in the UV/blue rest-frames indicating higher star formation rates at earlier times. Extensive work has been done on the UV LF which is largely driven by its importance as a tracer of star formation rate and assisted by increasing detection rates of distant Lyman break galaxies in deep surveys. We see $M_{UV}^\star$ brighten with increasing redshift and then turning over, thus roughly exhibiting the same behavior as the derived star formation history (Madau plot). The wide redshift range of available UV LF measurements makes it the only LF in which a non-monotonic evolution is distinctly seen in  $M^\star_{UV}$. In all other bands, the evolution of the Schecter parameters can be fitted with an analytic function to quantify the global evolution, while ``washing'' out outliers in the process. Several authors have parameterized the evolution in individual bands \citep[e.g.,][]{Lin99,Cirasuolo10}, but to our knowledge, our work is the first multi-wavelength parametric study of the evolution of the LF parameters. We find the following forms to fit the data well across our wide range of wavelengths and redshifts:
\begin{align}
  M^\star(z) & = M^\star(z_0) - 2.5\log{[(1+(z-z_0))^q]} \label{Msfit} \\
  \phi^\star(z) & = \phi^\star(z_0) \exp{\left[-p(z-z_0)\right]} \label{psfit}
\end{align}
and we assume the following {\it a priori} form for the faint-end slope
\begin{align}
  \alpha(z) & = \alpha(z_0) \left( z/z_0 \right)^r  \label{alfit}.
\end{align}
These fits are shown in Figure~\ref{fig_params}. 
For $M^\star(z)$ and $\phi^\star(z)$ we have taken $z_0$=0.8, but $z_0$= 0.01 for $\alpha(z)$. The other best-fit parameters are listed in Table~\ref{tab:bands}. Instead of selecting a preferred LF measurement for a given redshift in each band we have chosen to let all measurements contribute equally to the fitting process regardless of depth, area and sample size of the survey. Although there are a few notable discrepancies between the data and the fits we note that our IR-fluctuation results are unaffected as long as the fits remain good in the light shaded areas of Figure~\ref{fig_params}. These regions correspond to the distance for which the rest-frame emission is redshifted into the observed near-IR wavelengths of interest, defined to encompass the 1.25-4.5\mic\ range. In the following sections we will rely on lightcones extrapolated from the highest measured redshift, typically z$\sim$4, out to z$_{max}$=7 (see Table~\ref{tab:bands}). To account for the turnover observed in $M^\star_{UV}$, we only use our Equation (\ref{Msfit}) out to z$\sim$3 where they intersect the high-$z$ fitting formulae given by \citet{Bouwens11} which we adopt for z$\gtrsim$3.

Evolution is not easily discerned in the faint-end slope, $\alpha$, which by the very nature of surveys is hard to measure over large distances. For this reason we explore different scenarios for the behavior of $\alpha(z)$ which we explain in Section~\ref{sec:faintend}. In the \bL\ and \bM\ bands, the redshift range covered by the available measurements is so limited that we can only fit $M^\star(z)$ but not the other Schecther parameters. Thus, for these two bands we assume $\phi^\star(z)$ to take on the same form as the neighboring \bK-band. Fortunately, the data available in the \bL\bM-bands covers the redshift range of interest as is indicated by the shaded regions in Figure~\ref{fig_params}.

\begin{table}
\caption{Properties of the data shown in Figure~\ref{fig_params} and the best-fit evolution parameters of Equations (\ref{Msfit})-(\ref{alfit})}
\hfill
\begin{tabular}{ l c c c c c c }
  Band &  $\lambda_{eff}$ & $N$ & $z_{max}$ & $M^\star_0$,$q$ & $\phi^\star_0$,$p$ & $\alpha_0$,$r$   \\
   $(1)$    &  $(2)$           & $(3)$    &  $(4)$         &     $(5)$        &    $(6)$        &    $(7)$    \\
\hline
\hline

\bUV  &  0.15  &  24  &  8.0  &  -19.62,1.1  &  2.43,0.2  &  -1.00,0.086  \\
\bU  &  0.36  &  27  &  4.5  &  -20.20,1.0  &  5.46,0.5  &  -1.00,0.076  \\
\bB  &  0.45  &  44  &  4.5  &  -21.35,0.6  &  3.41,0.4  &  -1.00,0.055  \\
\bV  &  0.55  &  18  &  3.6  &  -22.13,0.5  &  2.42,0.5  &  -1.00,0.060  \\
\bR  &  0.65  &  25  &  3.0  &  -22.40,0.5  &  2.25,0.5  &  -1.00,0.070  \\
\bI  &  0.79  &  17  &  3.0  &  -22.80,0.4  &  2.05,0.4  &  -1.00,0.070  \\
\bz  &  0.91  &  7  &  2.9  &  -22.86,0.4  &  2.55,0.4  &  -1.00,0.060  \\
\bJ  &  1.27  &  15  &  3.2  &  -23.04,0.4  &  2.21,0.6  &  -1.00,0.035  \\
\bH  &  1.63  &  6  &  3.2  &  -23.41,0.5  &  1.91,0.8  &  -1.00,0.035  \\
\bK  &  2.20  &  38  &  3.8  &  -22.97,0.4  &  2.74,0.8  &  -1.00,0.035  \\
\bL  &  3.60  &  6  &  0.7  &  -22.40,0.2  & 3.29,0.8$^{*}$  &  -1.00,0.035  \\
\bM  &  4.50  &  6  &  0.7  &  -21.84,0.3  & 3.29,0.8$^{*}$  &  -1.00,0.035  \\

\hline
\end{tabular}
\tablecomments{ 1) Fiducial rest-frame band, (2) the effective wavelength in microns, (3) number of LFs used, (4) highest redshift of LF available in band, (5) Best-fit parameters for $M^\star(z)$ with $z_0$=0.8, (6) Best-fit parameters for $\phi^\star(z)$ with $z_0$=0.8 in units of $10^{-3}$$\mathrm{Mpc^{-3}}$, (7) The parameters for $\alpha(z)$ chosen to reflect the models (\HFE\&\LFE)  presented in Section~\ref{sec:faintend}. }
$^{*}${\footnotesize assumed to be the same as in $\widetilde{K}$}
\hfill
\label{tab:bands}
\end{table}



There is significant degeneracy in the Schechter parameters derived for a given galaxy population which can manifest itself in different values of ($\alpha$,$M^\star$,$\phi^\star$) depending on the LF-estimator used (see Appendix~\ref{sec:consistency}). The overall shape of the LF can appear similar despite different Schechter parameters typically resulting in a comparable value for the luminosity density, $\mathcal{L} = \phi^\star L^\star \Gamma (\alpha+2)$, which we display in the bottom panels in Figure~\ref{fig_params}. For example, \citet{Ilbert05} (VVDS) and \citet{Gabasch06} (FDF) derive comparable LFs depite giving very different values for the Schechter parameters. The general agreement of the $\mathcal{L}$-data and the curves, $\phi^\star(z) L^\star(z) \Gamma (\alpha(z)+2)$, indicates that our separate fits do not systematically over- or under-estimate the total luminosity density.

The second step is populating the lightcone seen from the standpoint of the observer. Light from distant galaxies appearing in the observed $X$-band was emitted at wavelength $\lambda_X/(1+z)$ i.e. at all rest-frame wavelengths shortwards of $\lambda_X$ throughout the redshift cone. We extract the Schechter parameters from our fits in Figure~\ref{fig_params} at the redshift defined by $z_i = \lambda_X / \lambda_{Y^i} - 1$ where $Y$ corresponds to our fiducial bands (\bUV\bU\bB\bV\bR\bI\bz\bJ\bH\bK\bL\bM) and $\lambda_Y < \lambda_X$. Our template LFs then become
\begin{equation} \label{eqn:poplight}
  \begin{split}
  \Phi_i(M|z_i) = 0.4 \ln{(10)}\phi ^{\star} (z_i) & \left(10^{0.4(M {\star} (z_i)-M)}\right)^{\alpha(z_i) + 1} \\
   & \times \exp{(-10^{0.4(M^{\star} (z_i)-M)})}.
   \end{split}
\end{equation}
The continuous evolution of the LF seen in the $X$-band is then obtained by interpolating the $\Phi_i$'s from $z=0$ to $z_{max}$. It should be noted that because of the $\alpha-M^\star$ degeneracy, our separated ($\alpha(z)$,$M^\star(z)$,$\phi^\star(z)$) fits used in Equation~(\ref{eqn:poplight}), cause some amount of deviation from the original shape of the LF. This is a small effect in comparison with the general disagreement between individual authors on the shape of the LF. We refer to Appendix~\ref{appendixA} where an independent method is used to populate the lightcone, in which the original shapes of the LFs are kept intact. We show that the two different methods produce the same results, confirming the validity of our standard treatment.

\begin{figure}
      \includegraphics[width=.49\textwidth]{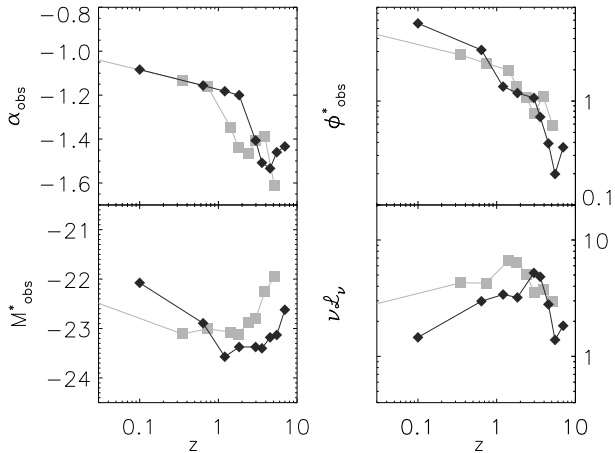}
      \caption{ Evolution of the Schechter parameters and the luminosity density seen in the observed at 2.2\mic\ (light squares) and 3.6\mic\ (dark diamonds). The values are extracted from the fits in \ref{fig_params} at the appropriate redshifts. $\phi^\star$ is in units of 10$^{-3}$Mpc$^{-3}$ and $\nu\mathcal{L}_\nu$ in units of 10$^{40}$erg$\cdot$s$^{-1}$$\cdot$Mpc$^{-3}$.  }
      \label{fig_obsparams}
\end{figure}

As an example we show in Figure~\ref{fig_obsparams} the Schechter parameters characterizing the LFs, probing the sky in two different observer-frame bands centered at 2.2, 3.6 \mic\ respectively.
Although the abundance of galaxies diminishes by itself at
high-$z$ according to our fits, we impose a limit
of $z_{max}\!=$7 in our modeling, beyond which we assume that ordinary galaxy
populations were not yet established. But due to the steep drop of $\phi^\star$ at high-$z$, our results are not
sensitive to this parameter: in fact, using $z_{max}=30$, yields results
nearly identical to our fiducial model. We emphasize that our evolution models are empirically supported out to z$\sim$4 only, beyond which we extrapolate the evolution deduced at lower redshifts.

In order to deduce the rest-frame LF from survey data, absolute magnitudes need to be derived from apparent magnitudes. We derive the flux from galaxies in our lightcone by backtracking the original procedure i.e. going from absolute magnitudes back to apparent magnitudes. This implies undoing any corrections the authors have made in this process
\begin{equation} \label{eqn:mapp}
  m = M + DM(z) + K(z) + E(z) + A_b(l,b),
\end{equation}
where $DM(z)$ is the distance modulus, $K(z)$ is the k-correction, $E(z)$ is the evolution correction and $A_b$ is the correction due to galactic extinction at the Galactic coordinate $(l,b)$. In LF measurements, authors typically use de-reddened magnitudes or correct for extinction using Galactic dust maps \citep{Schlegel98}. This correction can be large in the UV/optical but becomes less severe towards the infrared where we have $A_V/A_K$$\sim$7-10 approaching $\sim$15-20 in the IRAC bands \citep{Cardelli89,Indebetouw05}. We are only concerned with emission entering the Milky Way as near-IR where the extinction correction is typically well within 0.1 mag so we neglect it in Equation~(\ref{eqn:mapp}).  Correcting for evolution is intended to make a sample drawn from a distribution of redshifts reflect the true luminosity function at a given epoch (usually $z_{med}$ of the survey/bin) by accounting for changes in luminosity and number density over time \citep[e.g.,][]{Blanton03}. This has been done for some local surveys where a considerable spread in the redshift distribution leaves more cosmic time for evolution to take place. This typically results in corrections of $\sim$0.1 mag \citep{Bell03} but since the evolution correction simply acts to make the LF more accurate at a given redshift we do not need to make any adjustments.
The only magnitude adjustment in Equation~(\ref{eqn:mapp}) of concern is the k-correction \citep{Hogg02} which is needed to transform to the rest-frame by accounting for the redshifted SED of a given source. There are a variety of methods to deal with this SED dependence and we refer to Appendix~\ref{sec:consistency} for a more complete discussion of two commonly used treatment in the literature. From the k-corrected absolute magnitudes, we simply require the spectral independent term to account for the redshift into the observed frame, $K(z)=-2.5\log(1+z)$. Equation~(\ref{eqn:mapp}) is now reduced to
\begin{equation}
  m = M + DM(z) - 2.5\log(1+z),
\end{equation}
which is the conversion we use. In Section~\ref{sec:nc} we show that we recover the observed number counts to a very good accuracy using this methodology.

\subsection{The Faint-End LF Regime} \label{sec:faintend}

The source subtracted CIB fluctuations are isolated to faint sources. By the nature of galaxy surveys, the faint-end is generally poorly constrained causing large uncertainties and scatter in measurements of $\alpha$, especially at high-$z$. Because of this, many authors prefer to keep $\alpha$ fixed in their Schechter fits. Since the data does not show robust evolution in $\alpha$ in most bands (unlike $M^\star$ and $\phi^\star$) we explore variants of the behavior of the faint-end slope to get a feel for the sensitivity of CIB fluctuations to the abundance of faint galaxies. The substantial scatter in measurements of $\alpha$ leaves us some freedom in modifying the faint-end regime but we find that deep galaxy counts impose strict limits on the allowed range of faint-end slopes. This is most notable in \bB\bV\bR\bI, where a steep faint-end at $z$=1-3 leads to an overproduction of the observed $JHK$ number counts in the faintest magnitude bins (see Figure~\ref{fig_nc}). We therefore consider the range of allowed $\alpha(z)$ scenarios that collectively yield galaxy counts consistent with observations across all bands simultaneously. We leave $M^\star$ and $\phi^\star$ unchanged when varying $\alpha$ despite degeneracies in the parameters (see appendix \ref{appendixA}). We consider two models, {\it high faint-end} (\HFE) and {\it low faint-end} (\LFE), which, based on the resulting galaxy counts, are likely to bracket the true behavior of the faint-end of ordinary galaxy populations. These are shown in Figure~\ref{fig_params} and \ref{fig_nc} as the upper and lower boundaries of shaded regions. With the faint-end reasonably well constrained at z=0, ranging from -0.8 to -1.2, we fix $\alpha$ at these two values for \LFE\ and \HFE\ respectively and vary later evolution by changing the slope of the power-law in $\alpha(z)$ (called $r$ in Equation~(\ref{alfit}))\footnote{In the rest-frame UV/optical, where the low-$z$ contribution does not matter for the observed NIR, we fix the low-$z$ slope at -0.9 and -1.1 for \LFE\ and \HFE\ respectively.}. Our \HFE\ model is characteristic of strong steepening such as that found by \citet{Ilbert05} (VVDS) out to z$\sim$1 whereas the \LFE\ implies a more modest evolution, closer to that of \citet{Marchesini07,Marchesini12}. Our \LFE\ reflects a lack of evolution in the NIR i.e. $\alpha$$\sim$const., which seems to be favored by some authors \citep[e.g.,][]{Cirasuolo07}. We choose a faint-end cutoff for each template LF at $L_{cut}=10^{-4}L^\star$ for \LFE\ and $10^{-8}L^\star$ for \HFE, thereby extrapolating the LF to very low luminosities. For both scenarios we find $10^{-5}L^\star$ to be near saturation with flux contribution for fainter magnitude bins always being $<$0.02 \nW. Our ``default'' model is the average of \HFE\ and \LFE\ with a cutoff at $10^{-5}L^\star$.

We have chosen our \LFE/\HFE\ models so that they remain consistent with number counts data. The LFs dominating the faint counts in Figure~\ref{fig_nc} are mostly determined by the faint-end slope, $\alpha$, at high and intermediate redshifts and it is important to emphasize that more extreme faint-end evolution models generally yield number counts that are inconsistent with observations. Alternatively, one could in principle imagine an increase in the LF in the faintest magnitudes observed deviating from a Schechter function. In fact, such an upturn has been observed locally, for which a ``double'' Schechter function provides a better overall fit of the LF \citep[e.g.][]{Blanton05,Loveday12}. Allowing for a much steeper slope at z=0 to accommodate this possibility does not affect the resulting  CIB fluctuations because the surface density of sources on the sky tends to be dominated by populations at larger distances. This can be illustrated by examining the underlying LFs of the resulting galaxy number counts in Figure~\ref{fig_nc}, where the gray lines starting at the bright-end (from left) correspond to the local contribution (the thick line being the most local) moving to high-$z$ LFs to the right. The rapid redshift evolution of the cosmic volume element prevents a large surface density of low-$z$ sources and we find the faint counts always being dominated by populations at intermediate and high redshifts (z$\gtrsim$1). In order for low-$z$ sources to have sufficient densities to dominate the faint galaxy counts, and thereby also the unresolved fluctuations, we would need an extremely steep faint-end at z=0, becoming flatter towards increasing redshift i.e. $\alpha_{low-z}<\alpha_{high-z}$ which is the opposite of the observed evolution trend. Alternatively, a sudden upturn deviating from a Schechter form at faint magnitudes would need to shoot up by roughly two orders of magnitude. We therefore consider our \HFE\ scenario to be sufficiently extreme at low-$z$ and making it steeper does not have an effect on our results. On the other hand, if a significant upturn in the LF exists at z$>$0.5 (so far undetected), then this may result in a non-negligible contribution to the unresolved fluctuations. The large number of small halos predicted by the standard $\Lambda$CDM model permits such a scenario, especially if the first population of dwarfs with normal stellar populations formed in halos with mass $<$10$^9$~M$_\odot$ \citep{RicottiGS02a, RicottiGS02b, RicottiGS08}. For instance, if the ultra-faint dwarf galaxies recently discovered around the Milky Way can be identified as fossils of the first galaxies formed before reionization, that would imply that we have only discovered a small fraction of a widespread population of dwarfs which were almost certainly brighter in the past \citep{RicottiG05, Bovill09,Bovill11a,Bovill11b}. However, it is unclear how to make the flux from this population sufficiently large to reproduce the measured fluctuation signal and, furthermore satellite dwarfs are efficiently masked along with their host galaxy in fluctuation measurements as displayed by the masking typically having angular radius of $\simeq$15$^\prime$ (\citet{Arendt10}, see also Fig. A-3 in \citet{Kashlinsky12}). In this work we probe whether the known galaxy populations, which we extrapolate to faint magnitudes in our \HFE\ and \LFE\ limits, can account for the observed source-subtracted CIB fluctuations, and the question of the nature of the new populations that can explain these fluctuations is, while important, outside the scope of the current discussion.

\section{Number Counts and Background Light from LF Data} \label{sec:nc}


\begin{figure*}
      \includegraphics[width=0.99\textwidth]{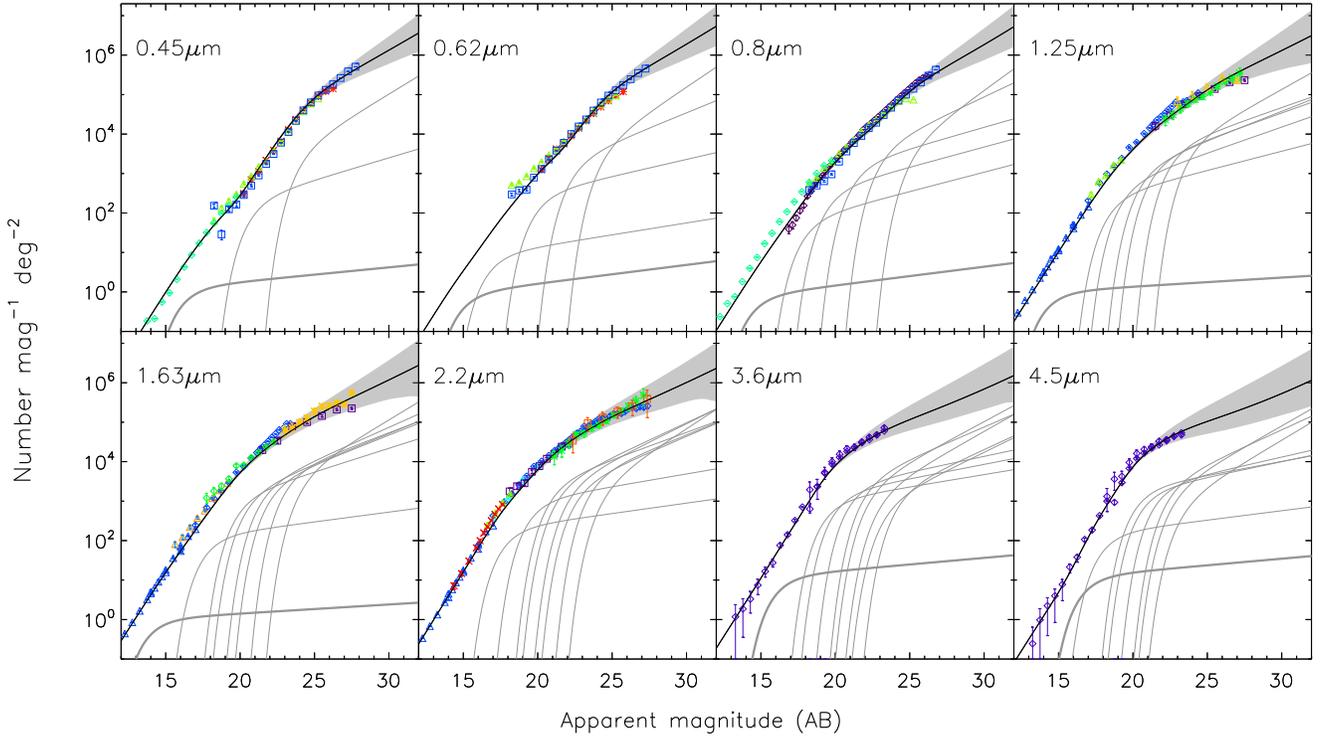}
      \caption{ Galaxy number counts in our default description (solid curve) including the regions of bracketed by our two extreme models, \HFE\ and \LFE\ (gray shaded areas). The gray curves show the underlying template LFs in our fiducial bands (Equation~(\ref{eqn:poplight})) which we interpolate and integrate to obtain the number counts via Equation~(\ref{eqn:nc}). The low-$z$ LF dominate the bright counts whereas high- and intermediate redshift LFs dominate the faint counts (from left to right). The most local available LF is shown as thick gray curves to demonstrate their negligible contribution to the faint counts. For 0.45-0.80\mic\ panels the data are from \citet{Capak04} (red asterisks),\citet{Capak07} (purple diamonds), \citet{McCracken03} (green triangles), \citet{Yasuda01} (turqoise diamonds) and \citet{Kashikawa04} (blue squares). Data in the 1.25-2.2\mic\ panels are taken from \citet{Vaisanen00} (green triangles), Dickinson et al. 1999 (purple squares), \citet{Maihara01} (green asterisks), \citet{Keenan10a} (blue triangles), \citet{Keenan10b} (blue diamonds)   \citet{Frith06} (yellow triangles), \citet{Thompson05} (yellow asterisks), \citet{Metcalfe06} (green diamonds), \citet{Quadri07} (turqiose triangles), \citet{Baker03} (purple squares), \citet{Minowa05} (orange squares), \citet{Huang97} (red crosses) and the 3.6-4.5\mic\ data comes from \citet{Fazio04} (purple symbols). }
      \label{fig_nc}
\end{figure*}
Galaxy number counts have the advantage of being free of the uncertainties associated with e.g. k-corrections and redshift determination which makes it an important test of both cosmology and galaxy evolution models. We project our lightcones onto the sky to obtain the galaxy number counts in each magnitude bin per unit solid angle:
\begin{equation}
  N(m) = \int \Phi(m|z)\frac{dV}{dzd\Omega}dz, 
\label{eqn:nc}
\end{equation}
where $dV/dzd\Omega$ is the comoving volume element per solid angle. In Figure~\ref{fig_nc} we display the number counts from Equation~(\ref{eqn:nc}) in the 0.45-4.5\mic\ range and compare with existing data in the literature. The good agreement between our modeling and observed counts demonstrates the validity of our method. We also display the range bracketed by out two limiting models for the faint-end slope of the LF, as discussed in Section~\ref{sec:faintend} (shaded areas). The gray curves in Figure~\ref{fig_nc} reflect the underlying template LFs contributing to the number counts in different redshift bins (bright/left to faint/right correspond roughly to low-$z$ to high-$z$), elucidating the different populations governing the source surface density on the sky. It is reassuring, although not surprising, that we recover the shape of the galaxy counts using independent observations (the only assumption being the Schechter parametrization of the LF). This explicitly confirms that our multi-wavelength collection of observed LFs provides an accurate description of the photometric properties of resolved galaxies on the sky.

\begin{figure*}
  \hfill
  \begin{minipage}[t]{.49\textwidth}
    \begin{center}
      \includegraphics[width=1.\textwidth]{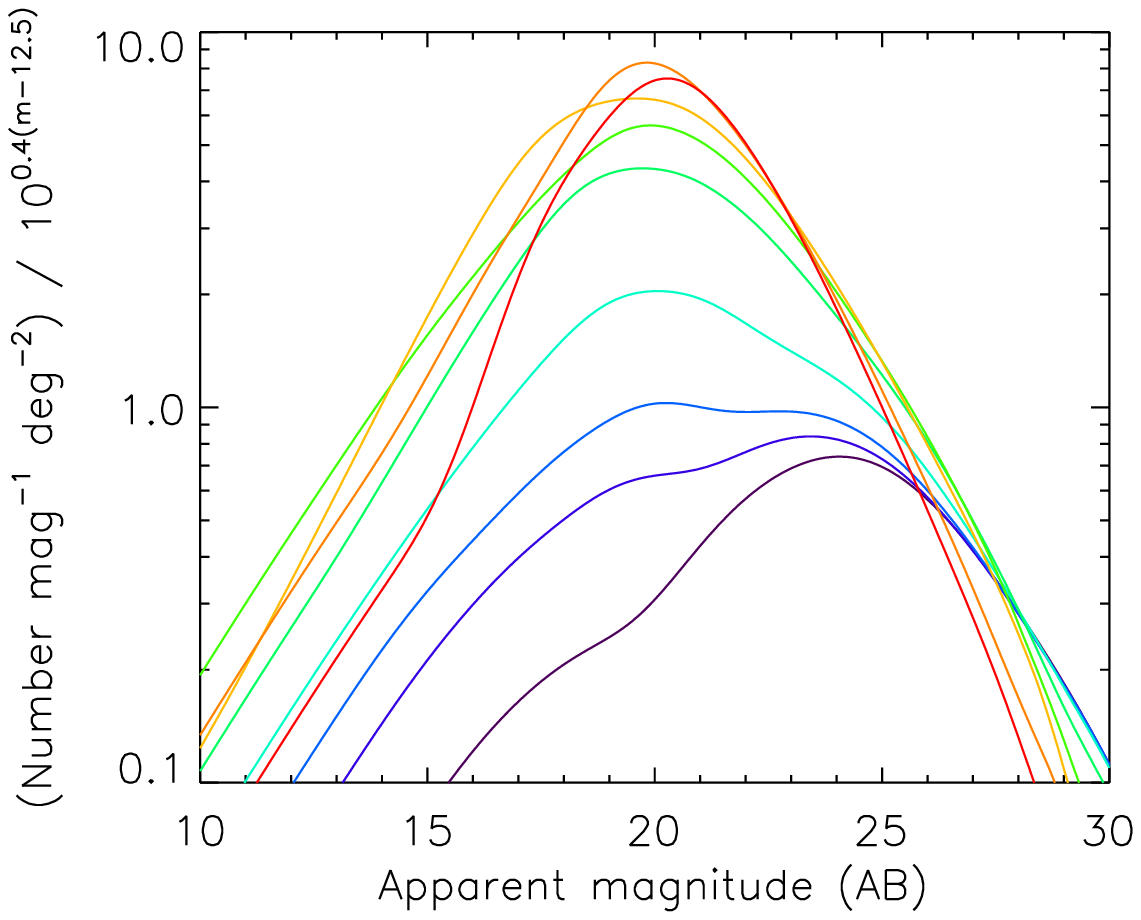}
    \end{center}
  \end{minipage}
  \hfill
  \begin{minipage}[t]{.49\textwidth}
    \begin{center}
      \includegraphics[width=1.\textwidth]{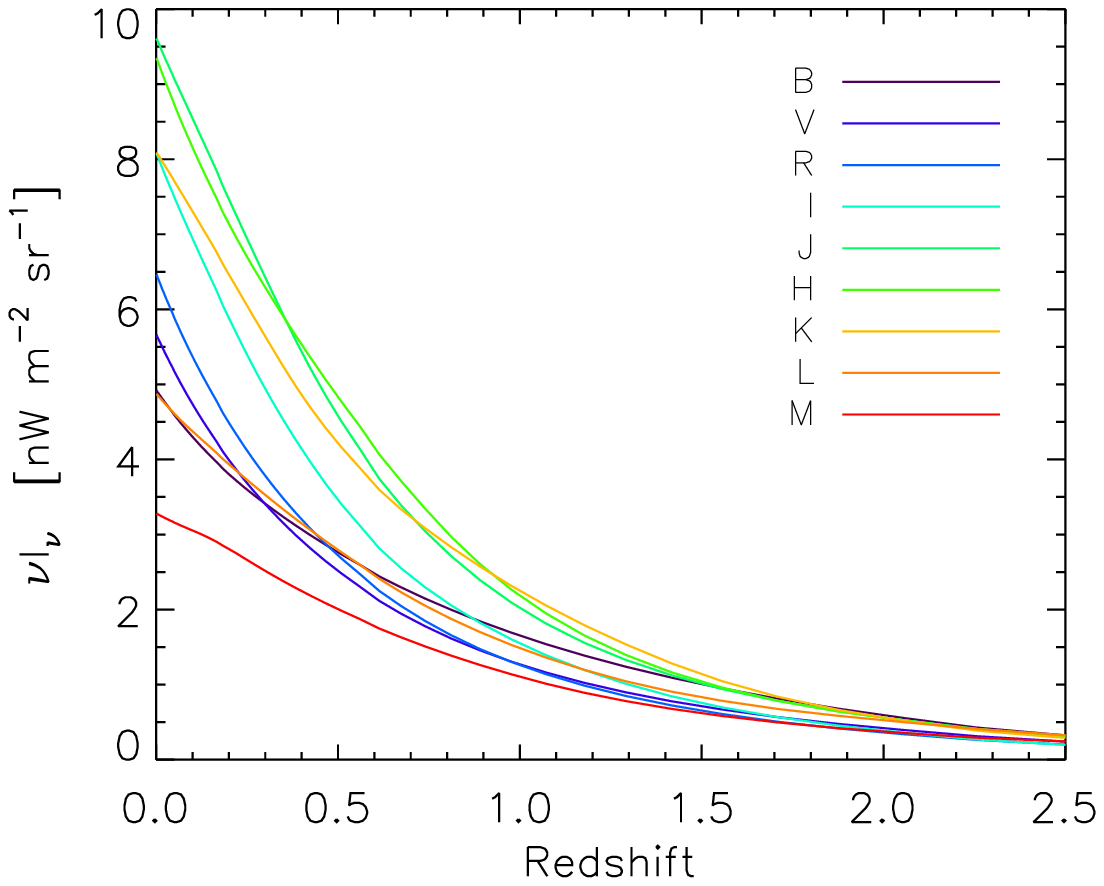}
    \end{center}
  \end{minipage}

      \caption{ {\it Left:} Our reconstructed number counts in BRIJKHLM compared across the spectrum. The counts have been multiplied by a slope of 10$^{-0.4}$ to bring out features in the shape. In this representation is proportional to the flux contribution from each magnitude bin. {\it Right:} The accumulation of integrated background light from galaxies over time. The flux builds-up from high-$z$ (right) to low-$z$ (left) reaching the present-day observed value listed in Table~\ref{tab:ebl} }
    \label{fig_ncmodels_eblz}
  \hfill
\end{figure*}

Figure~\ref{fig_ncmodels_eblz} (left) examines how the shape of the number counts varies across the spectrum (0.45-4.5\mic). Both the shape and amplitude of the counts are governed by the behavior of the ($\alpha(z)$,$M^\star(z)$,$\phi^\star(z)$)$_{obs}$-parameters shown in Figure~\ref{fig_obsparams} and some particular features deserve a few remarks. The bright counts all start out with a well known (Euclidian) slope of $d\log N/dm\!\sim$0.6 continuing down to $m$$\sim$18-20 where it flattens to $\sim$0.4. To first order this ``knee'' is simply caused by the transition from $M^\star$-dominated to $\alpha$-dominated regime. More specifically, a dip appears in the $BVRIJ$ number counts at $m\!\sim$18-20 which arises from the lack of very bright galaxies at higher redshifts, i.e. $M^\star_{obs}$ becomes fainter with redshift (see Fig.~\ref{fig_obsparams}). At higher redshifts (and shorter rest-frame wavelengths) we see a brightening again which is associated with star forming galaxies, bright in UV rest-frames. This brightening causes another feature at $\sim\! 25$~mag revealing a ``double-knee'' surrounding the dip. This is most pronounced in the $BVRI$-counts but disappears at longer observed wavelengths where the UV rest-frame becomes too distant. Beyond $m$$\sim$25-26, the counts are gradually diminished by the $\Lambda$CDM volume element. Depending on the exact faint-end model, the logarithmic slope in this regime is $\sim$0.2-0.3 in $BVRI$, decreasing as we go to longer wavelengths.

Another clear feature of the number counts seen in Figure~\ref{fig_ncmodels_eblz} (left), is the overall increase per magnitude bin as we go to longer observed wavelengths. We find the reasons for this to be twofold. First, the bright end is typically dominated by galaxies which are more luminous in the red bands such as the case of giant ellipticals. Therefore we see a larger number of them out to greater distances (in Fig.~\ref{fig_params} we clearly see $M^\star$ becoming overall brighter from blue to red). Second, when we look at the Universe through redder bands, we observe the redshifted light from bluer rest-frames emitted in epochs when the star formation activity was greater and consequently $M^\star$ was brighter. We further point out that our reconstructed counts are immune to confusion and agree well with the confusion corrected {\it Spitzer}/IRAC  counts of \citet{Fazio04} (confusion enters around $m_{AB}$$\sim$20-22).

We infer the amount of background light from galaxies from our reconstructed counts:
\begin{equation}
  \mathcal{F}_{\rm tot} \equiv \nu I_\nu =  \int  f(m)\frac{dN}{dm}dm,
\end{equation}
where $f(m)=\nu f_\nu$ of Equation~(\ref{eqn:AB}) and $\mathcal{F}$ is the integrated flux in units of \nW. Figure~\ref{fig_ncmodels_eblz} (right) shows how extragalactic background light builds up with cosmic time observed through $BVRIJKLM$. This results in present day values of the integrated background light of 9.6, 9.3, 8.1, 4.9 and  3.3 \nW\ at 1.25, 1.63, 2.2, 3.6 and 4.5\mic\ respectively  (see Table~\ref{tab:ebl}), which agree very well with Table 5 of \citet{Kashreview} and are also in general agreement with \citet{Madau&Pozzetti00}, but slightly lower than the values found by \citet{Keenan10b}. A subtle underestimation could be due to the smooth fitting of the LF evolution which smears out any abrupt variation of the Schechter parameters which could either be physical. The small deficit with respect to the EBL of \citet{Keenan10b} arises in the 21-23 mag range where we see a better agreement with \citet{Madau&Pozzetti00} and \citet{Maihara01}.

\section{Near-IR Fluctuations from Unresolved Galaxies} \label{sec:fluctuations}

 We now turn to evaluating the source-subtracted CIB fluctuations keeping in mind the procedure leading to their detection from raw images. If enough pixels remain in the maps after the masking of resolved sources, the fluctuations can be characterized via their angular power spectrum, which can then be computed more efficiently by using FFTs than the 2-point correlation function. For a detailed description of the process of reducing CIB fluctuation data in the {\it Spitzer}/IRAC analysis we refer to  \citet{Arendt10}.
\begin{table}[b]
\caption{Extragalactic Background Light}
\hfill
\begin{tabular}{ l c c c c c }
  Band & $m_{lim}$ & $m_{lim}$  & $m_{lim}$ & $m_{lim}$  & $\nu I_\nu$ \\
       &   $22$ & $24$  &  $26$ & $28$  & All  \\
\hline
\hline

$B$  &  3.33{\tiny$^{+1.72}_{-0.82}$} &2.26{\tiny$^{+1.56}_{-0.71}$} &1.17{\tiny$^{+1.24}_{-0.50}$} &0.52{\tiny$^{+0.88}_{-0.29}$} &4.92{\tiny$^{+1.81}_{-0.88}$} \\
$V$  &  2.95{\tiny$^{+1.54}_{-0.73}$} &1.90{\tiny$^{+1.36}_{-0.61}$} &0.96{\tiny$^{+1.05}_{-0.41}$} &0.42{\tiny$^{+0.73}_{-0.23}$} &5.65{\tiny$^{+1.73}_{-0.85}$} \\
$R$  &  2.86{\tiny$^{+1.54}_{-0.73}$} &1.75{\tiny$^{+1.31}_{-0.58}$} &0.85{\tiny$^{+0.98}_{-0.38}$} &0.37{\tiny$^{+0.67}_{-0.21}$} &6.56{\tiny$^{+1.82}_{-0.92}$} \\
$I$  &  2.81{\tiny$^{+1.58}_{-0.76}$} &1.58{\tiny$^{+1.27}_{-0.55}$} &0.72{\tiny$^{+0.92}_{-0.34}$} &0.30{\tiny$^{+0.61}_{-0.17}$} &7.97{\tiny$^{+2.01}_{-1.06}$} \\
$J$  &  2.59{\tiny$^{+1.56}_{-0.77}$} &1.20{\tiny$^{+1.10}_{-0.47}$} &0.48{\tiny$^{+0.72}_{-0.25}$} &0.18{\tiny$^{+0.45}_{-0.12}$} &9.60{\tiny$^{+2.40}_{-1.28}$} \\
$H$  &  2.25{\tiny$^{+1.50}_{-0.71}$} &0.96{\tiny$^{+0.96}_{-0.40}$} &0.36{\tiny$^{+0.57}_{-0.19}$} &0.13{\tiny$^{+0.34}_{-0.09}$} &9.34{\tiny$^{+2.59}_{-1.29}$} \\
$K$  &  1.74{\tiny$^{+1.41}_{-0.60}$} &0.69{\tiny$^{+0.82}_{-0.30}$} &0.24{\tiny$^{+0.44}_{-0.13}$} &0.08{\tiny$^{+0.23}_{-0.06}$} &8.09{\tiny$^{+2.52}_{-1.14}$} \\
$L$  &  0.98{\tiny$^{+1.05}_{-0.40}$} &0.34{\tiny$^{+0.57}_{-0.17}$} &0.11{\tiny$^{+0.27}_{-0.06}$} &0.03{\tiny$^{+0.12}_{-0.02}$} &4.87{\tiny$^{+1.72}_{-0.71}$} \\
$M$  &  0.75{\tiny$^{+0.83}_{-0.31}$} &0.24{\tiny$^{+0.45}_{-0.13}$} &0.07{\tiny$^{+0.20}_{-0.04}$} &0.02{\tiny$^{+0.09}_{-0.02}$} &3.28{\tiny$^{+1.21}_{-0.49}$} \\

\hline
\end{tabular}
\tablecomments{The upper and lower values are not error but correspond to the HFE/LFE evolution scenarios of the faint-end slope. All quantities are in \nW.}
\hfill
\label{tab:ebl}
\end{table}

The measured two-dimensional power spectrum from extragalactic sources consists of two components: i) the shot noise from the fluctuation in the number of unresolved sources entering the instrument beam, and ii) the clustering component arising from the correlation of galaxies on all scales. Additional power arising from local components such as Galactic cirrus and Zodiacal Light has been shown to be comfortably below the measured signal at 1-5\mic\ \citep{KAMM1,Matsumoto11,Kashlinsky12}. In comparing with observational data we adopt the convention for the power spectrum to approximate the root-mean-square fluctuations as $(q^2P_2(q)/2\pi)^{1/2}\!\sim\!\langle \delta F_{\mathbf{\theta}}^2\rangle^{1/2}$ \citep{Kashreview}. The angular power spectrum of galaxies projected onto the sky can be related to their evolving 3D power spectrum, $P_3(k)$, by the Limber approximation (for $\theta\ll$1 radian) which we adopt as modified by \citet{Fernandez10},
\begin{equation} \label{limber}
  P(q) = \frac{1}{c} \int \left[\frac{d\mathcal{F}}{dz}\right]^2 \frac{P_3(qd_A^{-1};z)}{ \frac{dt}{dz}d_A^2(z)} \frac{dz}{1+z},
\end{equation}
where $d_A$ is the comoving angular diameter distance. The quantity in the square brackets is the flux production rate which is {\it empirically} determined by our populated lightcones:
\begin{equation}  \label{eqn:flux}
  \frac{d\mathcal{F}}{dz} = \int_{m_{lim}}^\infty \!\! dm f(m)\frac{dN(m|z)}{dz}.
\end{equation}
\begin{figure}
      \includegraphics[width=0.47\textwidth]{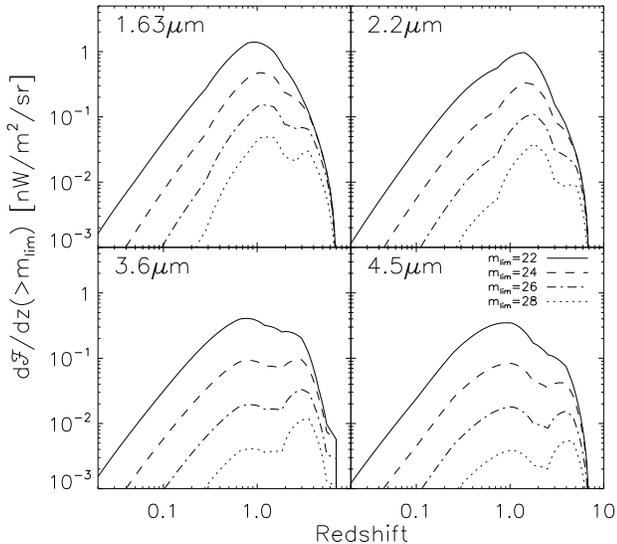}
      \caption{ Flux production rate (times z) as a function of redshift in the unresolved regime shown for limiting magnitudes of 22, 24, 26, 28 (solid, dashed, dot-dashed, dotted curves respectively). The total unresolved flux under each curve listed in Table~\ref{tab:ebl}. The figure illustrates how removal of ever fainter sources isolates the unresolved component to higher redshifts.}
      \label{fig_fzmag}
\end{figure}
It is important to note that the process developed in \citet{KAMM1,KAMM2} removes sources down to a fixed level of the shot-noise power (see Table~\ref{tab:sn}). This is equivalent to removing galaxies down to a limiting magnitude, $m_{lim}$, so that the remaining unresolved background is given by Equation~(\ref{eqn:flux}) integrated from $m_{lim}$ to $\infty$.
In Figure~\ref{fig_fzmag} we show the unresolved background from our modeling as a function of redshift, which illustrates the process of galaxy removal down to fainter magnitudes isolating the background to progressively higher redshifts. Note, that there is very little contribution ($\lesssim$0.1 \nW) from galaxies at z$\lesssim$1 after removing galaxies down to 26 AB mag. We find that for a limiting magnitude brighter than $\sim$24 mag, the unresolved flux is mostly dominated by $M^\star$ galaxies at intermediate redshifts whereas galaxies at the faint-end takes over once $m_{lim}$$\gtrsim$24. In Table~\ref{tab:ebl} we list the total integrated background in the 0.45-4.5\mic\ range including the unresolved background for different limiting magnitudes corresponding to the curves in Figure~\ref{fig_fzmag}.

\subsection{Shot Noise}

\begin{figure*}
  \begin{center}
      \includegraphics[width=0.8\textwidth]{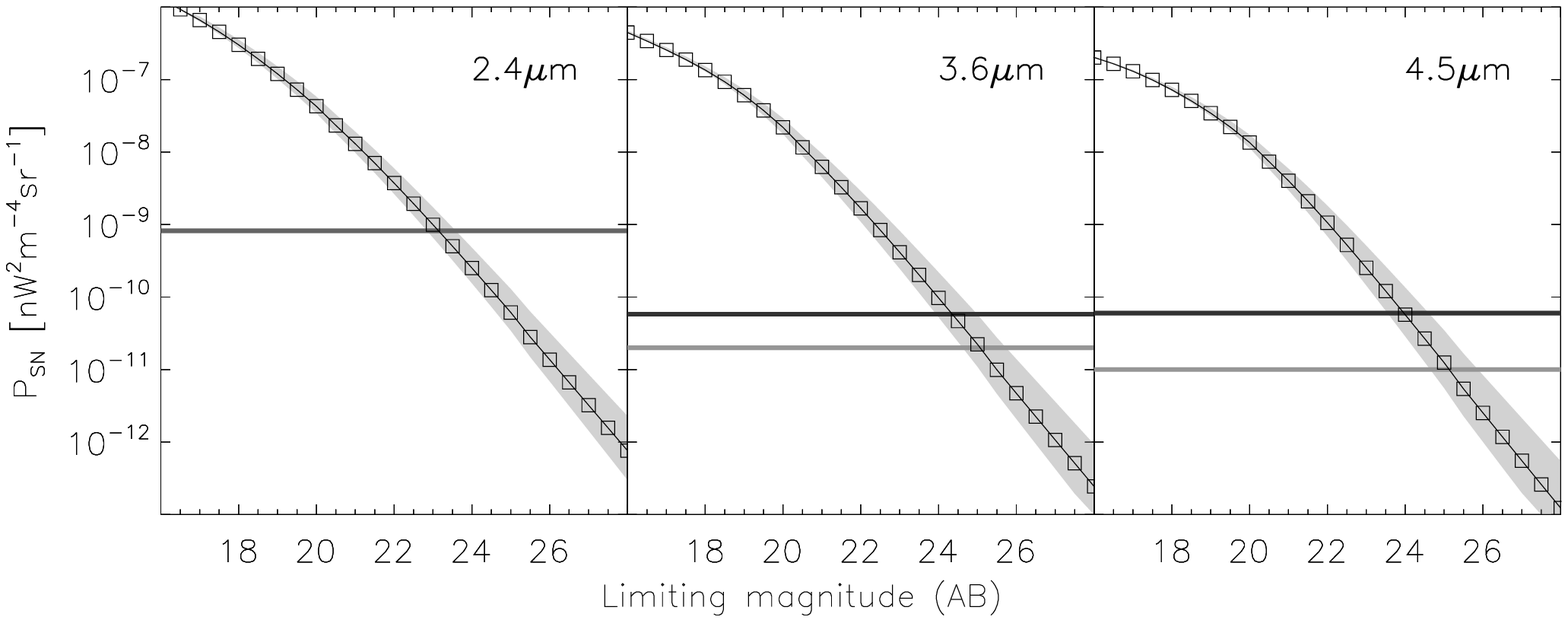}
      \caption{Shot-noise power amplitude after integrating the counts as a function of limiting magnitude (connected squares). The gray shaded area corresponds to the allowed range of the faint-end evolution of the LF. The thick gray lines show the levels of $P_{SN}$ reached by \citet{Matsumoto11} at 2.4\mic, \citet{KAMM1} (dark) and \citet{KAMM2} (light) at 3.6 and 4.5\mic. The intersection corresponds to the limiting magnitude reached in these studies. We tabulate these values in Table~\ref{tab:sn}. We point out that our model counts are immune to the effects of confusion. }
      \label{fig_snmag}
      \end{center}
\end{figure*}
The shot-noise level seen in fluctuation measurements is critically important in order to identify the nature of the unresolved populations \citet{KAMM3}. It can be described as statistical counting noise in the number of unresolved sources within the instrument beam and its power is,
\begin{equation}
  P_{SN} = \int \!\! dz \int_{m_{lim}}^\infty \! \! dm\hspace{2pt} f^2(m) \frac{dN(m|z)}{dz}.
\end{equation}
Shot noise is a directly measurable quantity and is not affected by confusion which may be present. This allows us to evaluate the effective limiting magnitude, $m_{lim}$, for a given shot noise level using our models which are also immune to confusion. We calculate the shot noise associated with galaxies in our lightcones and display it in Figure~\ref{fig_snmag} as a function of limiting magnitude at the relevant bands. As fainter galaxies are removed the shot noise drops steadily in the same manner as seen in measurements. At $\sim$22 mag we have already removed most $M^\star$ galaxies at z$\lesssim$1 beyond which the shot noise is mostly determined by the faint-end of the LF. The horizontal lines in Figure~\ref{fig_snmag} show the levels reached by the studies listed in Table~\ref{tab:sn}. The intersection with our models agrees well with \citet{KAMM2} claiming to have removed galaxies down to $m\sim$25-26 AB mag but is slightly brighter ($m\sim$24) for the levels reached by \citet{KAMM1} who claimed to reach $\sim$25 mag. Similarly, our shot noise levels agree well with those found by \citet{Matsumoto11} after removing galaxies down to AB magnitudes 22.9, 23.2 and 23.8 in the {\it AKARI}/IRC bands at 2.4, 3.2 and 4.1\mic\ respectively. Table~\ref{tab:sn} lists the limiting magnitude predicted for the shot noise levels reached in several studies.
\begin{table}
\caption{Limiting Magnitudes Implied by Shot Noise Levels}
\hfill
\begin{tabular}{ l c c }
  Reference  & $P_{SN}$ & $m_{lim}$ \\
   \hspace{20pt} Band         & {\tiny [10$^{-11}$ nW$^2$m$^{-4}$sr$^{-1}$]} & {\tiny (AB)} \\
\hline
\hline

\cite{Thompson07a} &  & \\
\hspace{20pt} F160W       &   $<$1.0       &  $\gtrsim$27\\ 
\cite{Thompson07b} &  & \\
\hspace{20pt} F110W       &   $<$1.8       &  $\gtrsim$27\\ 

\cite{KAMM1} &  & \\
\hspace{20pt} IRAC1$_{3.6\mu m}$       &   5.8       &  24.4{\tiny $^{+0.7}_{-0.5}$}    \\
\hspace{20pt} IRAC2$_{4.5\mu m}$     &    6.0     &     24.0{\tiny$^{+0.6}_{-0.4}$}     \\
\cite{KAMM2} &   & \\
\hspace{20pt} IRAC1$_{3.6\mu m}$    &    2.0    &      25.1{\tiny$^{+0.7}_{-0.4}$}      \\
\hspace{20pt} IRAC2$_{4.5\mu m}$    &    1.0      &    25.1{\tiny$^{+0.8}_{-0.5}$}      \\
\cite{Matsumoto11} && \\
\hspace{20pt} IRC$_{2.4\mu m}$      &    82$^*$   &      23.2{\tiny$^{+0.4}_{-0.3}$ }    \\
\hspace{20pt} IRC$_{3.2\mu m}$      &    33$^*$    &     23.3{\tiny$^{+0.5}_{-0.2}$ }      \\
\hspace{20pt} IRC$_{4.1\mu m}$      &    8.1$^*$    &    23.9{\tiny$^{+0.5}_{-0.3}$ }      \\

\hline
\end{tabular}
 \tablecomments{The upper and lower values are not error but correspond to the HFE/LFE evolution scenarios of the faint-end slope. $^*$The values are inferred from Figure 3 of \citet{Matsumoto11}.}
\hfill
\label{tab:sn}
\end{table}

We have defined $m_{lim}$ to separate resolved/removed galaxies from unresolved remaining sources. In practice, however, the accurate value of $m_{lim}$ reached depends on the source detection algorithm and the photometric aperture used to derive magnitudes. Furthermore, source extraction can become limited by confusion, depending on exposure and instrument beam. Since our underlying reconstruction of galaxy counts from LFs is immune to confusion, we assume that the measured shot noise levels serve as a reliable indicator for the faintest sources removed, $m_{lim}$. This obviously assumes that the source removal is done properly and does not introduce spurious signals in the background fluctuations as discussed at length in \citet{Arendt10}. It also assumes that the (quasi-)flat power seen on small scales is entirely due to shot noise dominating the contribution from non-linear clustering of galaxies which we discuss in the following subsection.

\subsection{Galaxy Clustering} \label{sec:clustering}

The shape and amplitude of the fluctuations produced in each redshift slice is dictated by the two evolving quantities in the Limber equation (eqn. \ref{limber}), i) the amount of light production given by our reconstructed $d\mathcal{F}/dz$ in a given band, and ii) the clustering pattern of the sources in this epoch, described by their three-dimensional power spectrum, $P_3(k,z)$. For the latter quantity we assume that on large scales sources cluster according to the observationally established concordance $\Lambda$CDM power spectrum. Prescriptions exist for non-linear evolution that modify the linear power spectrum in the regime where structures have collapsed out of the density field and linear theory breaks down \citep{Peacock96,Smith03}. However, luminous sources are known to be biased tracers of the dark matter distribution particularly in the non-linear regime where the correlations of sources depends on the {\it Halo Occupation Distribution} (HOD) of galaxies. We therefore consider a halo model description of the power spectrum which decomposes it into two terms, a two-halo term ($P^{2h}$) on large scales arising from the correlations of isolated halos, and a one-halo ($P^{1h}$) from correlations of particles within the same halo on small scales (e.g. \cite{Ma&Fry00}). We follow the treatment of \citet{Cooray&Sheth02} and write,
\begin{equation}
  P^{gal}(k) = P^{1h}(k) + P^{2h}(k),
\end{equation}
where,
\begin{align} \label{eqn:halo_model}
  P^{1h}(k) &= \int dM \frac{dn}{dM} \frac{2\langle N_{sat} \rangle \langle N_{cen} \rangle u(k|M) + \langle N_{sat} \rangle ^2 u^2(k|M)  }{\bar{n}_{gal}^2},  \\
  P^{2h}(k) &= P^{lin}(k)  \left[ \int dM \frac{dn}{dM} \frac{\langle N_{gal} \rangle}{\bar{n}_{gal}}b(M) u(k|M) \right]^2,
\end{align}
and $dn/dM$ is the halo mass function \citep[from]{Sheth&Tormen01}, $\bar{n}_{gal}$ is the average number density of galaxies, $P^{lin}(k)$ is the linear $\Lambda$CDM power spectrum (computed using the transfer function of \citet{Bardeen86}), $u(k|M)$ is the normalized Fourier transform of the halo profile \citep{NFW96}, and $b(M)$ is the halo bias \citep[from][]{Sheth&Tormen01}. The occupation number has been separated into central galaxies, $\langle N_{cen} \rangle$, and satellite galaxies, $\langle N_{sat} \rangle$, such that
\begin{equation}
  \langle N_{gal} \rangle = \langle N_{cen} \rangle + \langle N_{sat} \rangle.
\end{equation}
We take the mass dependence of our HOD model to follow the four parameter description of \citet{Zheng05}:
\begin{align}
  \langle N_{cen} \rangle  &= \frac{1}{2}\left[ 1 + {\rm erf}\left( \frac{\log M - \log M_{min} }{\sigma_{\log M}}\right) \right], \\
  \langle N_{sat} \rangle  &= \frac{1}{2}\left[ 1 + {\rm erf}\left( \frac{\log M - \log 2M_{min} }{\sigma_{\log M}}\right) \right]  \left(\frac{M}{M_{sat}} \right)^{\alpha_s},
\end{align}
where $\langle N_{cen} \rangle$ is characterized by $M_{min}$, the minimum halo mass that can host a central galaxy and $\sigma_{\log M}$, which controls the width of the transition of the step from zero to one central galaxy. The satellite term has a cut-off mass which is twice as large as the one for central galaxies and grows as a power-law with a slope of $\alpha_s$, normalized by $M_{sat}$. This form has been explored both numerically and observationally. Since the measurements of HOD-parameters are obtained from samples of resolved galaxies at low-$z$, their validity may not extend into the unresolved regime or, in particular, to higher redshifts. Since we are concerned with the unresolved regime it is important to note that the measured cut-off mass of central galaxies, $M_{min}$, is typically set by the lowest luminosity probed by the survey so halos may continue to host central galaxies to lower masses but are excluded due to selection criteria. In Section~\ref{sec:nc} we showed how the unresolved light is typically dominated by the faint-end of the LF for $m$$\gtrsim$25 with most bright central galaxies removed out to $z$$\sim$3 in measurements of CIB fluctuations. One would also expect the masking to eliminate most of the surrounding satellite galaxies. We have adopted the following parameters of the HOD-model motivated by SDSS measurements of \citet{Zehavi11}: $\sigma_{\log M}\!=0.2$, $M_{min}\! =\! 10^{9}M_\odot$, $M_{sat}\! =\! 5\cdot10^{10}M_\odot$, and $\alpha_s \! = \! 1$ where we have deliberately chosen a lower cut-off reflecting low mass halos hosting galaxies well into the unresolved regime, and a lower $M_{sat}$ allowing for large amounts of unresolved satellite galaxies, while keeping $\alpha_s \! = \! 1$. It should be noted, that in the absence of any HOD-assumptions, a simple linear $\Lambda$CDM clustering with typical bias, $b^2P^{lin}(k)$, produces nearly identical fluctuations on large scales. The one-halo term has white-noise power spectrum ($P=$const) with its amplitude limited from above by the measurements at small scales and so its modeling is irrelevant to interpreting the clustering signal at scales $\ga 1^\prime$.

We assume that unresolved sources in our lightcones are uniformly mapped onto the halo distribution i.e. the clustering is independent of galaxy luminosity. In practice however, we expect the most luminous galaxies to be removed in the masking process along with most of the accompanying satellites. This could motivate one to introduce an upper mass limit in the integrals in Equations~(\ref{eqn:halo_model}), $M_{max}(z)$. However, this would require an additional mass-to-light ratio assumption and since it would always result in a decrease of the clustering amplitude, we do not apply $M_{max}(z)$ and consider the result to be an upper limit for the resultant power spectrum. This includes the mass-dependent bias which is similarly integrated over the entire range of occupied halos ($\gtrsim$10$^9$). The large scale (linear regime) galaxy bias seen by \citet{Zehavi11} in the local SDSS sample is $b \approx$1 when all galaxies are included. At somewhat higher redshifts, \citet{Granett12} find $b=1.38 \pm 0.05$ averaged over 0.5$<$z$<$1.2. Further increase of the linear bias with redshift is expected on theoretical grounds as collapsing density peaks were increasingly rare in the past. The bias prescription used here shows the same general behavior \citep{Sheth&Tormen01}. Several CIB studies at far-IR wavelengths claim a linear bias as high as $b$=2-3 for far-IR sources \citep[e.g.,][]{Lagache07,Viero09} but at these redshifts, the samples are already biased towards the most luminous objects due to selection effects. If anything, we expect the bias to be lower in the faint and unresolved regime after the more strongly biased luminous galaxies are masked and removed.

The large scale fluctuations are always dominated by clustering in the linear regime (two-halo term). On the other hand, the non-linear clustering described by the one-halo term in Equation~(\ref{eqn:halo_model}) exhibits a $P(k)$=const behavior ($\delta F$$\propto$$q$) making it indistinguishable from shot noise in measurements. Given that we found excellent agreement between the shot noise in our models and the measurements at the same magnitude levels, there does not seem to be any need to invoke non-linear clustering to explain fluctuations on small-scales (unless the data points deviate from a simple white noise spectrum $\delta F$$\propto$$q$). In addition, we explored the pure dark-matter treatment of the non-linear clustering of \citet{Smith03} but find it to be inconsequential in comparison with the shot noise dominated fluctuations on small scales.  Although we see the one-halo term contributing somewhat to the \HFE\ fluctuations in Figure~\ref{fig_fluctuations}, it becomes less relevant if one accounts for the more massive halos being masked/removed. In fact, we will see in Section \ref{sec:results} that current fluctuation measurements place a limit on the amount of non-linear power in the unresolved regime.

\subsection{Comparison with fluctuations from the Millennium Simulation and Semi-Analytic Models}

To compare our results with the clustering of halos seen in large scale N-body simulations, we have made use of the theoretical lightcones constructed by \citet{Henriques12}. These mock catalogs are based on semi-analytical models for galaxy evolution \citep{Guo11} which are implemented on two very large dark matter simulations, the Millennium Simulation \citep{Springel05} and the Millennium-II Simulation \citep{Boylan-Kolchin09}. The simulations provide a description of the evolving spatial distribution of dark matter halos and subhalos whereas the nature of the baryonic content is described by the latest version of the semi-analytical Munich model \citep{Guo11}. The Millennium Simulation follows structure formation in a box of side 500$h^{-1}$Mpc comoving with a resolution limit of $\sim$$10^{10}$$h^{-1}$M$_\odot$ whereas the Millennium-II Simulation focuses on a region of 100$h^{-1}$Mpc but with complete merger trees down to $\sim$$10^{8}$$h^{-1}$M$_\odot$\footnote{The Millennium Simulation and the resulting lightcones of \citet{Henriques12} assume a WMAP1-based cosmology \citep{Spergel03} with parameters $h$= 0.73, $\Omega_m$=0.25, $\Omega_\Lambda$=0.75, $n$=1 and $\sigma_8$=0.9 which are slightly different that our adopted parameters of $h$= 0.7, $\Omega_m$=0.3, $\Omega_\Lambda$=0.7 but this is of no appreciable consequence for the results in Figure~\ref{fig_fluctuations}.}. \citet{Henriques12} use the Millennium Simulation only in their study, limiting the faint-end of the LF to halos $>$10$^{10}$$h^{-1}$M$_\odot$. Even so, the predicted faint near-IR counts are higher than observations suggest due to an unusually high abundance of relatively low mass galaxies ($\sim$10$^{10}$M$_\odot$) at z$>$1 (\citet{Guo11} tuned their model to match the local populations). A comparison of the predicted correlation function of these models with local SDSS data shows decent agreement for massive galaxies whereas correlations of low mass systems are overpredicted, particularly at small separations.  \citet{Henriques12} also neglect the effects of dust and PAH emission and consider only starlight using stellar synthesis models of \citet{Bruzual&Charlot03} and \citet{Maraston05}.  For a detailed description of these models we refer to \citet{Springel05}, \citet{Guo11} and \citet{Henriques12}.

Despite the limitations mentioned above, we find that this study provides a useful comparison to our fluctuation analysis. After constructing images using the publicly available mock data of \citet{Henriques12}, we calculate the projected angular power spectrum, convolved with the instrument beam. We analyze two independent regions observed in H, K, IRAC1 and IRAC2 each covering 1.4$\times$1.4 degrees on the sky. We extract all galaxies in the magnitude range $m_{lim}$$<$$m$$<$30 to produce the unresolved fluctuations which we display alongside our results in Figure~\ref{fig_fluctuations}. Because of the overabundance of faint galaxies at 3.6 and 4.5\mic\ in the semi-analytical description of \citet{Guo11}, we need to remove galaxies down to 0.2 mag deeper than the $m_{lim}$ listed in the panels in order to normalize to a common shot noise level. This NIR overabundance (despite the resolution limit of $\sim$10$^{10}$$h^{-1}$M$_\odot$) results in the Millennium fluctuations (dark-gray shades in Figure~\ref{fig_fluctuations}) being in closer agreement with our \HFE\ scenario at 3.6 and 4.5\mic\ but can otherwise be considered to be consistent with our main results.

\subsection{Results} \label{sec:results}

\begin{figure*}[!t]
\begin{center}
      \includegraphics[width=0.95\textwidth]{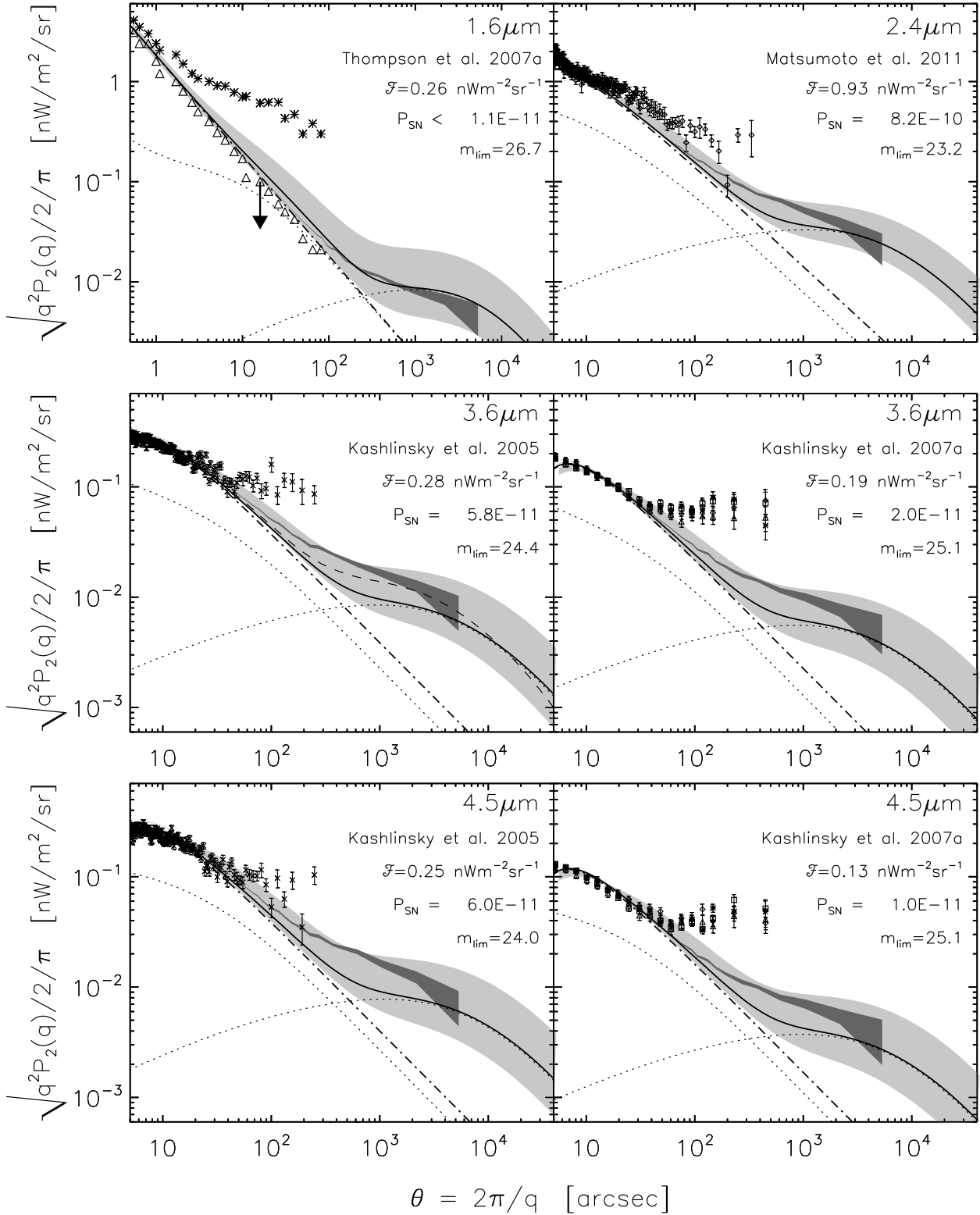}
      \caption{ Models of the unresolved near-IR fluctuations compared to measurements from authors listed in the panels. We have chosen the limiting magnitude such that the models are normalized to the shot noise levels reached in these studies (including a contribution from a one-halo term). The solid curves show the total contribution from clustering and shot noise whereas the light shaded areas indicate the region bracketed by our \HFE\ and \LFE\ models. These are all suppressed by the instrument beam on small scales. The dotted lines indicate the separate one-halo and two-halo terms of the power spectrum. Shown in each panel is the total unresolved flux associated with the default model ($\mathcal{F}$), the values of $P_{SN}$ (in units of nW$^2$m$^{-4}$sr$^{-1}$) and the associated $m_{lim}$. The dark shaded regions correspond to fluctuations arising from galaxies in the lightcones of \cite{Henriques12} derived from the Millennium Simulation in the magnitude range $m_{lim}$$<$$m$$<$30. Because of their overabundance of faint galaxies at 3.6 and 4.5\mic\, we have increased the $m_{lim}$ of the Millennium fluctuations by 0.2 mag to normalize to the correct shot noise levels. In the 3.6\mic\ panel we also show the default model from \citet{Sullivan07} (dashed line). In the 1.6\mic\ panel the notation follows Fig. 2 of \citet{Thompson07a}: asterisks correspond to fluctuations with all sources removed whereas the triangles indicate their estimate of the instrumental Gaussian noise.}
      \label{fig_fluctuations}
\end{center}
\end{figure*}

\begin{figure*}
  \begin{center}
      \includegraphics[width=0.8\textwidth]{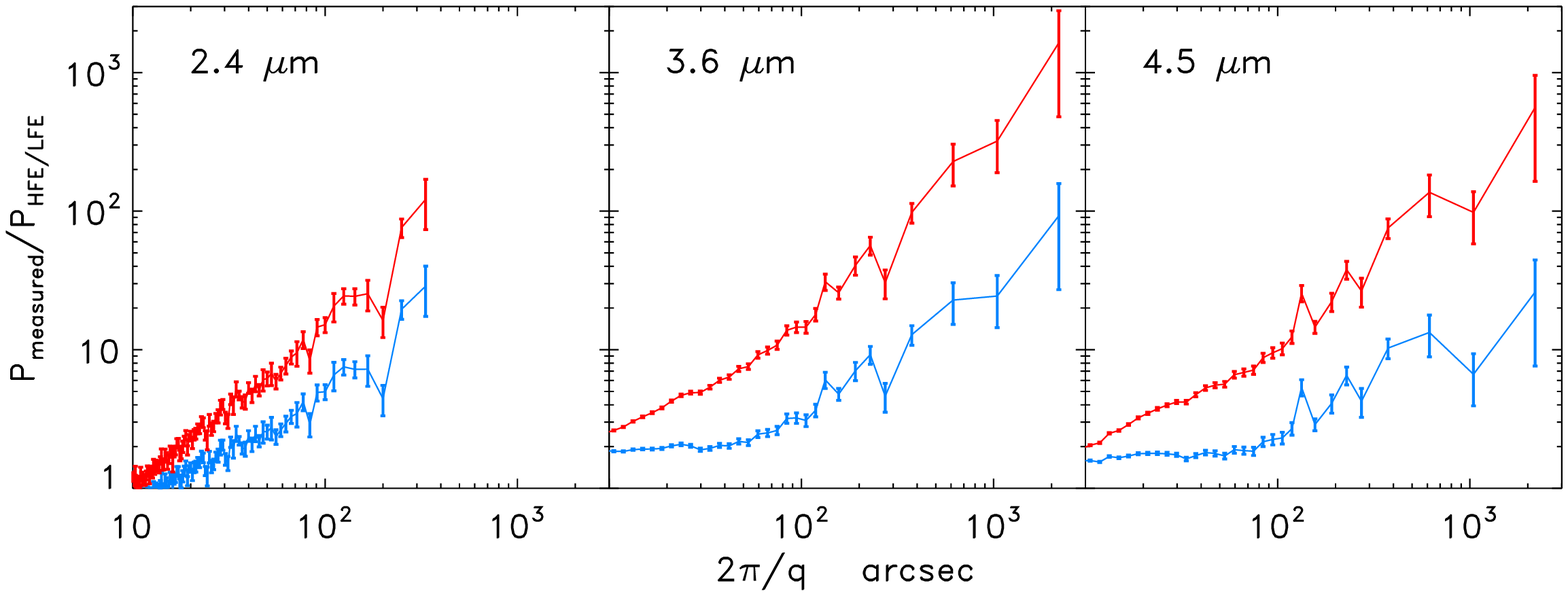}
      \caption{The lines show the ratio of the measured source-subtracted power spectrum from the {\it AKARI} 2.4\mic\ data and the latest {\it Spitzer}-based measurements at 3.6 and 4.5\mic\ \citep{Kashlinsky12} to the \HFE\ and \LFE\ expectations (red/upper and blue/lower respectively). The results show that the measured CIB fluctuations continue to diverge from our models as we go to larger scales and are thus unlikely to result from extra-biasing of these faint populations: in order to explain the measured signal the biasing would have to be 1) scale-dependent, i.e. non-linear, 2) biasing amplification would have to be more non-linear on scales where the amplitude of the underlying correlation function is weaker (larger scales), and 3) the biasing would have to be different at 3.6 and 4.5\mic.}
      \label{fig_seds}
\end{center}
\end{figure*}

With the emission history reconstructed from LFs and the sources distributed according to the halo model in Section~\ref{sec:clustering}, we projected onto the sky the clustering pattern in our NIR lightcones using Equation~(\ref{limber}) and display the results in Figure~\ref{fig_fluctuations}. The limiting magnitudes have been chosen such as to normalize the shot noise (dot-dashed lines) to the measurements shown in each band. The shot-noise is seen to dominate the fluctuations on small scales whereas the clustering component becomes significant at arcminute scales.
In the display we have chosen to focus on 1.6, 2.4, 3.6 and 4.5\mic\ where we can compare with measurements from {\it Hubble}/NICMOS, {\it AKARI}/IRC, and {\it Spitzer}/IRAC. Our models have been convolved with the beam profile (or PSF) of these instruments. It is immediately clear from Figure~\ref{fig_fluctuations} that the contribution from known galaxy populations falls short of the measured clustering signal in every band shown. We briefly discuss each comparison:

\citet{KAMM2} find excess fluctuations of $\delta$$F$$\sim$0.05-0.1 \nW at arcminute scales in the {\it Spitzer}/IRAC channels after removing sources down to $\sim$25 mag or shot-noise levels $P_{\rm SN} \! \lesssim  \! 3\times 10^{-11}$ nW$^2$/m$^4$/sr. It can be seen from Figure~\ref{fig_fluctuations} that the known sources remaining at the measured shot-noise levels cannot account for the observed fluctuations for any faint-end modeling of the LF. We have displayed the data of \citet{KAMM1} and \citet{KAMM2} in panels side-by-side illustrating that the discrepancy gets larger as galaxies are removed to deeper levels. The unresolved flux associated with our default model is 0.18 \nW\ in the deepest 3.6\mic\ maps of \citet{KAMM2}, so in order to explain the observed level of the excess fluctuations the relative levels of the source-subtracted CIB fluctuations would have to be close to non-linear, $\delta \! F/F \! \sim$1, all the way to $\sim \! 10^\prime$. The spatial spectra of the CIB fluctuations from the known galaxy populations is such that the gap increases toward large scales if this behavior of the source-subtracted CIB fluctuations continues as observed (say, $\sim \! 1^\circ$), so these fluctuations would have to be in the same (quasi)non-linear regime at much larger scales making it more difficult to explain them with the known galaxies. The additional linear biasing to amplify the arcminute scale signal to the observed levels but this would require $b\! \sim \! 6-20$ which is highly unlikely for small systems in the $1 \! \lesssim \! z  \! \lesssim \! 3$ range where most of the flux is produced. This, however, can be shown not to be viable in light of the latest Spizter-based results submitted after this paper. \citet{Kashlinsky12} measure the source-subtracted CIB fluctuations on sub-degree angular scales confirming the earlier results and identifying, for the first time, the fluctuations spectrum to $\sim 1^\circ$ where the discrepancy continues to grow. Figure~\ref{fig_seds} shows the ratio of the measured power spectrum from the new large scale {\it Spitzer}/IRAC data of \citet{Kashlinsky12} to the power spectra of our \HFE\ and \LFE\ (red and blue), illustrating that the data keeps diverging from our models out to $\sim 0.5^\circ$. This shows that if one were to model the measured CIB fluctuations with extra biasing of the known galaxy populations, the biasing would have to be 1) highly scale-dependent, i.e. more prominent on larger scales, where density pattern is in linear, 2) the resultant biasing factors would have to be huge reaching amplifications of over two order of magnitude at the largest scales, and 3) the biasing would have to be wavelength dependent attesting to the different discrepancy ratios in each band. This argues further against the detected CIB fluctuations arising from the a faint-end extension of the known populations.

Following the HOD-model in Section~\ref{sec:clustering} we find that small scale power from the one-halo term ($P^{1h}$) is lower than the shot noise ($P_{SN}$). For the \HFE\ model however, the two are comparable at 3.6 and 4.5\mic\ which leads to a slight increase in the upper values of $m_{lim}$ in Table~\ref{tab:sn} (by 0.3 mag) as normalized by the measured shot noise levels. The fact that our shot noise, calculated from number counts, is in good agreement with the small scale data points of \citet{KAMM1,KAMM2}, argues against a significant clustering on small scales. It must therefore remain at or below the measured level of $P_{SN}$, otherwise sources would have to be removed to $m_{lim}\! \gtrsim \! 26$, which is not possible in the deepest {\it Spitzer}/IRAC maps.

\begin{figure*}[t]
  \begin{center}
      \includegraphics[width=0.85\textwidth]{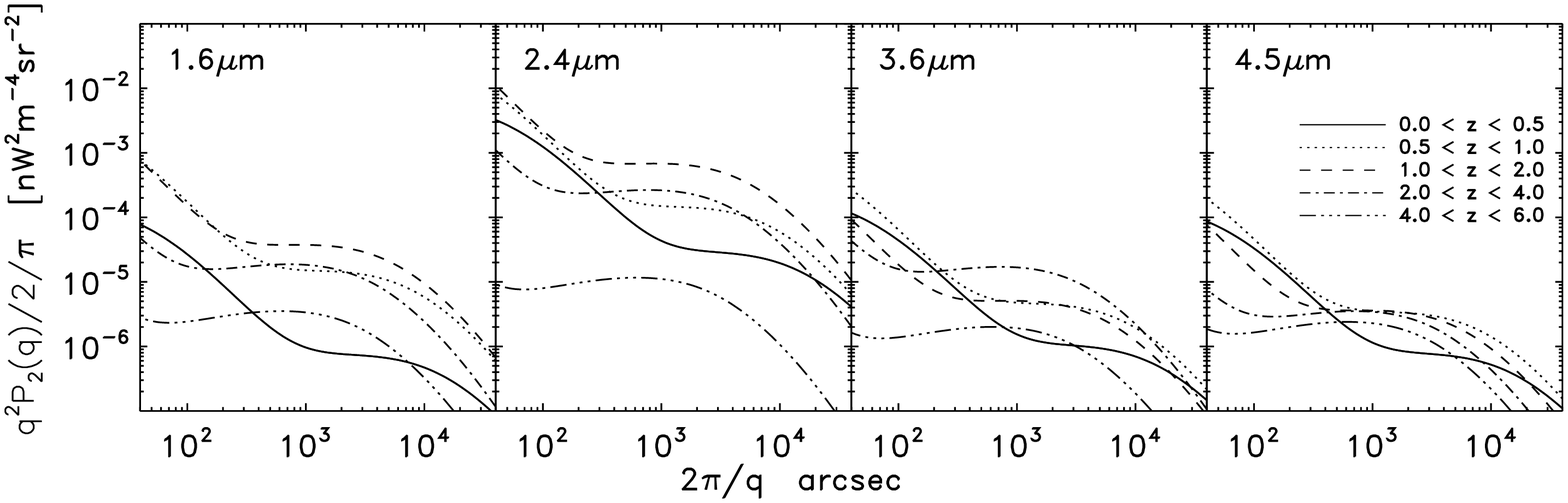}
      \caption{The contribution of different redshift bins to the unresolved IR-fluctuations shown in Figure~\ref{fig_fluctuations} for the $m_{lim}$ indicated in the panels. The 3.6 and 4.5\mic\ panels correspond to the models at the shot noise levels of \citet{KAMM2}. The different set of lines correspond to the redshift bins indicated in the legend. This illustrates that depending on the observed band and the depth of source removal, the unresolved fluctuations from known galaxies are dominated by populations at different epochs. The amplitude and shape is governed by 1) the flux production history (see Fig.~\ref{fig_fzmag}), and 2) the evolving power spectrum, $P_3(k,z)$. The non-linear clustering component is important at low-$z$ but moves towards small scales for higher $z$. The dependence on the comoving angular diameter distance, $d_A(z)$ (see Eqn. (\ref{limber})) is easily seen as the peak of the $\Lambda$CDM power spectrum shifts towards smaller scales with increasing redshift. }
      \label{fig_slice}
\end{center}
\end{figure*}
\cite{Chary08} stack deep {\it Spitzer} exposures to detect faint ACS galaxies beyond the detection threshold of the frames used in \citet{KAMM2} and explore the sensitivity of the IR-fluctuations to these ACS sources. Their stacked source detections down to 26.0-26.2 mag imply a net flux of 0.12-0.35 \nW. For comparison, the flux associated with our lightcones in the 25-26.2 mag range is 0.04 and 0.2 \nW\ at 3.6\mic\ for \LFE\ and \HFE\ respectively with 0.04-0.35 \nW\ from still fainter galaxies, $>$26.2 mag. We note that \citet{KAMM4} demonstrate observationally the negligible correlations on arcminute scales  between the source-subtracted CIB maps, as constructed by their self-calibration procedure \citet{Arendt10} and ACS source maps.

\citet{Thompson07a} measure fluctuations at 1.6\mic\ on scales out to 80$^{\prime \prime}$ using {\small {\it HST}/NICMOS} (and at 1.1\mic\ in \citet{Thompson07b}) and ascribe the signal to faint galaxies emitting at redshifts $z$$\sim$0.5-1.5. Their fluctuations at 80$^{\prime \prime}$ have amplitudes of $\sim$0.4 \nW, which is a factor of 2-7 times higher than the total unresolved component, 0.06-0.20 \nW, for sources fainter than $>$28 mag, indicating that the clustering of the underlying galaxies must be highly non-linear. For their CIB fluctuation levels to be reconciled with our empirical estimates, the one-halo term would have to be significantly higher, but then its amplitude would overshoot the data at all the other NIR wavelengths. If we take the upper limit on the shot noise at these wavelengths to be at the levels of the estimated instrument noise of \citet{Thompson07a}, then our shot noise already matches at AB magnitude of $\sim$27 (see triangles in Figure~\ref{fig_fluctuations}). But even at that level we cannot reproduce the fluctuations (asterisks) with the clustering of known galaxy populations out to $1^\prime$. \footnote{We note that in the context of \citet{Thompson07a}, our theoretical magnitude limit, $m_{lim}$ at 1.6\mic\ should no longer be taken as a definitive boundary between resolved and unresolved sources because ACS images at shorter wavelengths were used to remove sources which can translate to a wider spread in magnitudes at 1.6\mic\ (due to different exposures and different SEDs of individual sources).} We point out in this context the clearly visible outer halos of the sources removed by Thompson et al (2007a, see their Fig. 4) whose contribution to their CIB fluctuations shown may be significant and should be estimated for more quantitative conclusions at 1.6 \mic.

The \citet{Matsumoto11} measured fluctuations at 2.4, 3.2, 4.1\mic\ using data {\it AKARI} satellite and conclude that they are consistent with stars from early epochs confirming the identification proposed in \citet{KAMM1}. The left panel in Fig. \ref{fig_seds} confirms that the AKARI signal at 2.4\mic\ cannot be explained by the remaining known galaxy populations.

In Figure~\ref{fig_fluctuations} we also display the default model from \citet{Sullivan07} (dashed lines) who combined a halo model and conditional luminosity functions to calculate IR-fluctuations at 3.6\mic. Our models have a somewhat lower amplitude considering the fact that we use $m_{lim}$=24.4 as opposed to the 25.3 mag used by \citet{Sullivan07} (and quoted in \citet{KAMM1}) but the two are in rough agreement. For $m_{lim}$=25.3 our unresolved flux is 0.1 \nW\ (\LFE) which is roughly consistent with the 0.08 \nW\ found by \citet{Sullivan07}. However, they claim that the fluctuations measured by \citet{KAMM1} at 3.6\mic\ can be explained by galaxies in the magnitude range 25.3 to 28.8 (AB) at $z$$\sim$1-3. This is a somewhat puzzling conclusion when comparing their model with the data in Figure~\ref{fig_fluctuations} as it clearly fails to account for the clustering excess\footnote{The data-points from \citet{KAMM1} appear only in the electronic version of \citet{Sullivan07}.}.

In Figure~\ref{fig_slice} we show the contribution of different redshift bins to the unresolved IR-fluctuations for the $m_{lim}$ indicated in the panels of Figure~\ref{fig_fluctuations}. This illusrates the different epochs in which unresolved galaxy populations contribute to the fluctuations in different observed NIR bands. The redshift dependence  is governed by 1) the flux production history (see Fig.~\ref{fig_fzmag}), and 2) the evolving power spectrum, $P_3(k,z)$. The Figure also reflects the dependence on the comoving angular diameter distance, $d_A(z)$ (see Eqn. (\ref{limber})) with the overall clustering pattern shifting towards smaller scales with increasing redshift.

\section{Summary and Discussion} \label{sec:discussion}

We have reconstructed the emission histories seen in the near-IR of present-day observers to model the unresolved CIB fluctuations and compared with current measurements. Our compilation of \NLF\ luminosity functions used to populate lightcones at $z$$<$7 reproduces the observed number counts remarkably well and accounts for the features shaping them. We assume the Schechter-type LF and model the evolution of its parameters from the available datasets. We then considered high and low faint-end LF limits within the constraints permitted by deep galaxy counts data. Extending these to faint magnitudes and to high-$z$ we calculated the range of unresolved background flux in deep images and derived CIB-fluctuations from these galaxy populations predicted by the standard $\Lambda$CDM clustering power spectrum. We find good agreement between the predictions of our analysis and semi-analytical galaxy evolution models combined with the large scale Millennium N-body simulation.

By varying the limiting magnitude of source subtraction we normalize our models to the observed shot noise levels, finding good agreement with the depths reached in current fluctuation measurements. We show that the known galaxy populations fail to account for the observed source subtracted CIB clustering signal in either \LFE\ or \HFE\ limits. Although, in principle, by varying $m_{\rm lim}$ one can find a population of brighter galaxies that matches the measured clustering amplitude at some fiducial angular scale, the associated shot noise levels always imply that all such populations have been removed in the source subtraction thereby not contributing to the unresolved fluctuations. Thus it means that the emitters producing the source-subtracted CIB fluctuations on arcminute scales are below the detection limits of current surveys and furthermore, cannot be a part of the known evolving galaxy populations. In other words, the only way to reproduce the clustering excess with extragalactic sources is by introducing a new population of sources that are significantly fainter than the detection threshold of current instruments i.e., a highly clustered population with low shot noise.

The high isotropy of the CIB fluctuation signal measured in Spitzer IRAC data \citet{Kashlinsky12} argues strongly against the signal originating in Galactic or Solar system foreground emissions as well as very local extragalactic sources. Since the observed galaxy populations (extrapolated to very faint limits) cannot explain the measurements, the CIB fluctuations must originate in new populations so far unobserved in galaxy surveys. \citet{KAMM4} show that there are no correlations between the ACS maps with sources down to AB mag of $\simeq$28, and the source-subtracted CIB maps from \citet{KAMM2}. This implies that either the CIB fluctuations originate in a large unknown population of very small systems at low/intermediate redshifts, or they are produced by high redshift, z$\gtrsim$7, populations whose Lyman break (at rest 0.12\mic) is shifted passed the longest ACS channel (at 0.9\mic).


\section*{ACKNOWLEDGMENTS}

This work was supported by NASA Headquarters under the NASA Earth and Space Sciences Fellowship Program - Grant NNX11AO05H. KH is also grateful to The Leifur Eiriksson Foundation for its support. MR acknowledges partial support by NASA Grant NNX10AH10G and NSF CMMI1125285. We also thank to B. Henriques, R. Keenan and T. Matsumoto for useful exchanges and data. Our compiled database of Schechter parameters is available upon request.

\bibliographystyle{apj}
\bibliography{myrefs}

\appendix

\section{A. LF Binning and Interpolations} \label{appendixA}

\begin{figure*}[t]
      \includegraphics[width=0.97\textwidth]{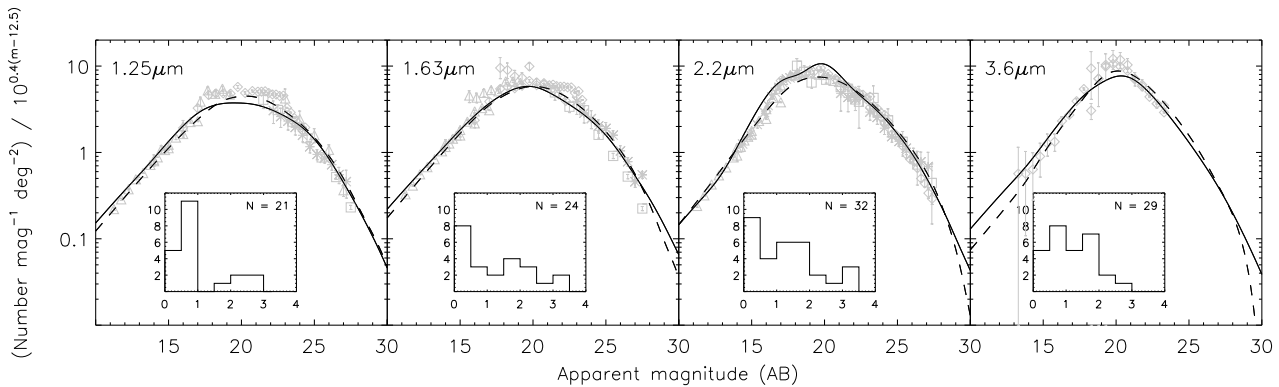}
      \label{fig_functionwise}
      \caption{Comparison between our default method (dashed) and the alternative method presented here (solid). The two curves agree to within 20\% in the range shown. The data shown in the background is the same as in Figure~\ref{fig_nc}. The insets show the redshift distribution of LFs avalable in for the calculation in each band (i.e. $\lambda_0$$-$$\Delta\lambda<\lambda^{obs}<\lambda_0$$+$$\Delta\lambda$).
 }
\end{figure*}

Because of degeneracy in ($\alpha$,$M^\star$,$\phi^\star$), different sets of Schechter parameters can represent LFs of very similar shapes. The method used in Section~\ref{sec:poplight} disentangles the Schechter parameters to fit their evolution individually. In addition to this, we used an alternative approach in which the shape of each measured LF is kept intact. We took each LF in its rest-frame and redshift the associated emission to the observed wavelength, $\lambda^{obs}=\lambda^{rest}(1+z)$. We examine the all LFs that meet the criterion $\lambda_0$$-$$\Delta \lambda<\lambda^{obs}<\lambda_0$$+$$\Delta \lambda$ where $\lambda_0$ is the center of the NIR band and $\Delta \lambda$ is roughly the {\small FWHM} of the filter. The inserts in Figure~10 show the redshift distribution of available LFs which can be observed through JHKL. In a given band, we place each LF in redshift bins and take the functional average of $\Phi(M)$ in common bins so that we have a unique LF at each redshift. We thus have template LFs, $\Phi_i(M|z_i)$, in each of the observed NIR bands and the rest of the analysis is identical to that in Section~\ref{sec:poplight} following from Equation~(\ref{eqn:poplight}) (we interpolate the evolution and project the populations onto the sky). The major shortcoming of this method is the redshift information. Averaging over several LF in a common redshift bins is immune to the effects of Schechter parametrization but comes at the cost of crude evolution i.e. the sampling of $z$ is determined by the number of $z$-bins. As seen in Figure~\ref{fig_functionwise} there is no guarantee that there exists a LF measurement falling into $\lambda_0$$-$$\Delta\lambda<\lambda^{obs}<\lambda_0$$+$$\Delta\lambda$ in each redshift bin. In this case we borrow LFs from neighboring wavelengths scaling them according to synthetic spectra. Figure~10 shows that despite these limitations, we obtain very comparable number counts to the ones in Section~\ref{sec:nc}, agreeing to within 20\% in the relevant magnitude range.

\section{B. Consistency notes}  \label{sec:consistency}

\subsubsection{K-correction}

Calculating the absolute magnitudes of a galaxy sample requires a k-correction to account for the offset in the rest-frame and the observed SED due to the cosmological redshift (e.g. \cite{Hogg02,Blanton03b})
\begin{equation} \label{Mabs}
  M_X = m_X - DM(z) - K(\lambda_X,\lambda_{X^\prime})
\end{equation}
where $X$ refers the band of interest. The k-correction can be written (in AB magnitudes)
\begin{equation}
  K(z) = (m_{X^\prime}-m_X) -2.5\log_{10} (1+z)
\end{equation}
where $m_X$ is the observed brightness of a galaxy at redshift $z$ and $m_X^\prime$ is its rest-frame brightness in $X$-band. The exact value of the k-correction requires knowledge of the spectral energy distribution (SED) of the source and is commonly evaluated by assuming a template SEDs based on the galaxy type/color. This treatment is fairly reliable for low-$z$ galaxies but the correction can become large for high-$z$ galaxies and dominate the uncertainty in the derived LF, especially in the blue bands. Recent multiband photometric surveys offer a robust way of reducing this SED dependency by utilizing magnitudes in multiple bands to constrain the best-fit SED . Not only does multiband coverage indicate SED shape but when probing the LF in the rest-frame band $Y$ centered at $\lambda_Y$, the galaxy flux can be sampled in the band $X$ which is closest to $\lambda_Y(1+z)$. In other words, the observed filter ($X$) that best matches the redshifted rest-frame band of interest is the one that minimizes $|\lambda_X - \lambda_Y(1+z)|$. The k-correction needed then becomes the matter of setting this quantity to exactly zero which is typically a small correction. We can rewrite Equation~(\ref{Mabs}) in this framework
\begin{equation}
  M_Y = m_X - DM(z) - K(\lambda_X,\lambda_Y(1+z))
\end{equation}
where the SED dependence of the k-correction is now small even at high redshifts. Backtracking the original procedure to apparent magnitudes now requires simply $K(z)=-2.5\log_{10}(1+z)$ which we use in Equation~\ref{eqn:mapp}.

\subsubsection{Photometric Systems}

Unfortunately, there is no photometric system which is universally accepted and the different ways used to evaluate the apparent magnitude of galaxies in the survey can introduce biases affecting the derived luminosity functions (see \citet{Bessel05} for a review of photometric systems). As the flux from a galaxy diminishes from the center it will eventually drop below the background noise to be missed by the aperture. Photometric systems based on total magnitudes, such as S\'ersic, are usually preferred since they directly quantify the physical flux while apertures such as Kron and Petrosian will always suffer from missed light to some extent. However, total magnitudes typically assume an extrapolated profile which is model-dependent and has larger measurement errors \citep{Cole01}. The Petrosian system can be advantageous since it compensates for the effects of seeing by increasing the fraction of the light recovered from a galaxy when its angular size is small \citep{Blanton01}.  Despite this, Petrosian magnitudes are found to underestimate S\'ersic by 0.2 mag \citep{Strauss02,Blanton01}. Likewise, 2MASS Kron and isophotal magnitudes may account for only 50-80\% of the total flux in the most extreme cases \citep{Andreon02}). For example, \citet{Smith09} show that their UKIDSS Petrosian magnitudes can be up to 0.5 mag fainter than 2MASS Kron magnitudes. The fraction of the lost flux increases towards fainter galaxies and may cause a systematic underestimation of the faint-end luminosities as well as the luminosity density. \citet{Hill11} provide a good analysis of the effects of different photometric systems used in surveys. They find an overdensity of faint galaxies when compared with the best-fit Schechter function irrespective of the aperture system used and show that a Schechter function parametrization does not provide a good fit at the faint-end. They also show that the use of a photometric systems based on total magnitudes (e.g. S\'ersic extrapolated) have a systematically steeper faint-end slope than photometric systems based on Kron or Petrosian magnitudes. They further show that the r-band Kron \& Petrosian photometry underestimates the luminosity density by at least $\sim$15\% as they do not account for missing light. \citet{Blanton03} show that the difference of the luminosity density resulting from Petrosian and S\'ersic magnitudes should be within $<$0.1 mag in the SDSS bands and not worth correcting for given the limitations of both systems.  Still many authors apply a correction to estimate the total magnitudes in order to derive quantities such as the luminosity density in physical units (e.g. \cite{Kochanek01,Bell03,Eke05}). These can be as high as 0.3 mag in the K-band. It seems that uncertainties in the LF may be dominated by the aperture governing the fraction of flux recovered, especially at the faint end.

\subsubsection{Luminosity Function Estimators}

In this paper we use LFs derived from a variety of different LF estimators. The choice of LF estimator is unlikely to be a major source of discrepancy between the LFs derived by different authors although it can lead to different combinations of the Schechter parameters. The most commonly used methods are i) the $1/V_{max}$ method \citep{Schmidt68}, ii) the Sandage-Tammann-Yahil maximum likelihood method (STY) \citep{Sandage79} and iii) the StepWise maximum Likelihood Method (SWLM) \citep{Efstathiou88}. The $1/V_{max}$ method is reliable in the sense that it simultaneously gives the shape and normalization of the LF requiring no assumption on the parametric form for the LF. However, it suffers from systematic biases in the presence of density inhomogeneities in the observed field. The STY method is typically preferred when estimating the LF over multiple fields since it has been shown to be unbiased to large scale structure and does not require binning of the data \citep{Efstathiou88}. It  does however require an assumption of a functional form of the luminosity function. The SWML method is widely used since it makes no assumption of the LF shape while still being insensitive to large scale structure. \citet{Willmer97} compare the properties of each LF estimator and show how different LF estimators tend to be biased towards the faint-end either overestimating or underestimating the slope, depending on the estimator and the underlying catalog. In order to minimize such effects one routinely compares the outputs of more than one method (e.g. \cite{Bouwens07,Ilbert05,Cirasuolo10}).



\label{lastpage}
\end{document}